\documentstyle[referee,epsfig]{l-aa}
\def\simlt{\ \raise -2.truept\hbox{\rlap{\hbox{$\sim$}}\raise5.truept   %MC
\hbox{$<$}\ }}                                                          %
\def\simgt{\ \raise -2.truept\hbox{\rlap{\hbox{$\sim$}}\raise5.truept   %
\hbox{$>$}\ }}

\def\cm2s{{\rm cm}^{-2} ~{\rm s}^{-1}}
\def\ergcm2s{{\rm erg} ~{\rm cm}^{-2}~ {\rm s}^{-1}}

\def\be{\begin{equation}}
\def\ee{\end{equation}}
\def\arcmin{$'$}

\begin{document}

   \thesaurus{03     % A&A Section 3: extragalactic
              ( 11.03.4; %{\bf Galaxies: clusters: individual:} $\ldots$
                11.09.3; %intergalactic medium
                12.03.3; %Cosmology: observations
%               12.04.1; %dark matter
                11.01.2; %Galaxies: active
                13.25.2)} % X-rays: galaxies
%               13.25.3)} % X-rays: general

\title{First gamma-rays from galaxy clusters. Preliminary evidence of
the association of galaxy clusters with
EGRET unidentified gamma-ray sources}

%   \subtitle{First gamma-rays from galaxy clusters}

   \author{S. Colafrancesco
}

   \offprints{S. Colafrancesco}

   \institute{INAF - Osservatorio Astronomico di Roma,
              via Frascati 33, I-00040 Monteporzio, Italy.
              Email: cola@coma.mporzio.astro.it
}

\date{received: June 12, 2001; accepted: February 28, 2002}
%\markboth{S. Colafrancesco}{First gamma-rays from galaxy clusters}
%\authorrunning {Sergio Colafrancesco}
%\titlerunning {First gamma-rays from galaxy clusters}

\maketitle

\markboth{S. Colafrancesco}{First Gamma-rays from galaxy clusters}

\begin{abstract}
The vast majority of the celestial gamma-ray sources detected so far have not yet been identified with secure
counterparts at other wavelenghts. Here we report the preliminary evidence of a probable association between
galaxy clusters and unidentified gamma-ray sources of high galactic latitude ($|b|>20$ $\deg$) in the Third EGRET
catalog. All of the clusters which are most probably associated with EGRET sources show evidence of strong radio
emission either because they host radio galaxies/sources in their environments or because they have a radio halo
or relic inhabiting their intracluster medium. The cluster radio emission suggests that the relativistic particles
(electrons, protons), which are diffusing in the intracluster medium, might be also responsible for their
gamma-ray emission. Beyond the spatial associations of clusters with unidentified EGRET sources, we found a
correlation between the radio flux at $1.4$ GHz of the cluster's brightest source and the gamma-ray flux, $F(>100
~{\rm MeV})$, of the associated EGRET source. For the most probable EGRET-cluster associations we also found a
further correlation between the X-ray luminosity of galaxy clusters and the gamma-ray luminosity of the associated
gamma-ray source under the hypothesis that the EGRET sources have the same cluster redshifts. Such correlations
are consistent with the theoretical expectations and strengthen the probability of a true, physical association
between galaxy clusters and gamma-ray sources.


\end{abstract}

\section{Introduction}
The large part of the gamma-ray sources detected with the EGRET instrument (Kanbach et al. 1988)  on board the
CGRO satellite have not yet been identified with secure counterparts at other wavelenghts because of  the poor
spatial resolution of the EGRET instrument. In fact, $170$ gamma ray sources out of the 271 found in the Third
EGRET catalogue are not yet identified with firmly established counterpart (Hartman et al. 1999). Most of the
unidentified gamma-ray sources are found at low galactic latitudes, $|b| \simlt 20$ $\deg$, and are likely to
belong to our Galaxy (Gehrels et al. 2000). Fifty of these sources are found at high galactic latitudes, $|b|
\simgt 20$ $\deg$, and there are several hints that they are of extra-galactic nature (Grenier 2001). Among the
identified extra-galactic gamma-ray sources observed with EGRET, most of them are AGNs (Hartman et al. 1999) but
there are no firm evidence that the remaining unidentified EGRET sources can be associated with another population
of active galaxies. In fact, most of the unidentified EGRET sources have a rather low flux variability while AGNs
usually show a strong flux variability in the gamma-rays (Urry \& Padovani 1995; Ulrich et al. 1997).
\\
 Galaxy clusters are bright sources of X-rays produced through bremsstrahlung emission from a hot (with
temperature $T \sim 10^7 - 10^8 $ K), optically thin (with number density $n \sim 10^{-3}$ cm$^{-3}$), highly
ionized intracluster (hereafter IC) gas (mainly consisting of a population of thermal electrons and protons) in
nearly hydrostatic equilibrium with the overall gravitational potential of the structure (see, e.g., Sarazin 1988
for a review). Many galaxy clusters also show the presence of non-thermal emission phenomena like extended radio
halos (see, e.g., Giovannini \& Feretti 2000), likely produced by synchrotron emission of relativistic electrons
either accelerated in the intracluster medium (hereafter ICM) by merging shocks or produced in the decay of dark
matter annihilation products (see, e.g., Colafrancesco 2001a, Colafrancesco \& Mele 2001). Many clusters also host
bright radio (or active) galaxies living in their environment. These active galaxies can inject relativistic
particles into the ICM through the interaction of radio jets with the surrounding medium (Blandford 2001).
 The presence of relativistic
particles in the ICM has been also suggested to explain the emission excesses observed in some clusters in the EUV
(Lieu et al. 1999, Bowyer 2000) and in the hard X-rays (Fusco-Femiano et al. 1999-2000, Rephaeli et al. 1999,
Kaastra et al. 1999, Henriksen 2000). However, there is no evidence in the EGRET database for a detection of
gamma-ray emission in the direction of a few selected clusters like Coma (Sreekumar et al. 1996) and Virgo.\\
 There are, nonetheless, several
theoretical motivations to expect that galaxy clusters can indeed be extended sources of gamma-rays emitted in the
decay of neutral pions, produced either in the interaction of cosmic ray protons with the ICM protons ($p p \to X
+ \pi^0 \to \gamma + \gamma$; see Colafrancesco \& Blasi 1998, Volk \& Atoyan 2000) or in the annihilation of dark
matter particles ($\chi \chi \to X + \pi^0 \to \gamma + \gamma$; see Colafrancesco \& Mele, 2001). The secondary
electrons produced in the previous mechanisms (see, e.g., Blasi \& Colafrancesco 1999; Colafrancesco \& Mele 2001)
can also produce additional gamma-ray emission through both bremsstrahlung and Inverse Compton Scattering (ICS)
against the Cosmic Microwave Background (CMB) photons. Also primary cosmic ray electrons can produce a diffuse
flux of gamma-rays due to non-thermal bremsstrahlung (see Sreekumar et al. 1996, Colafrancesco 2001b) and ICS of
the CMB photons. On top of such diffuse emission, the gamma-ray emission emerging from individual `normal'
galaxies (Berezinsky et al. 1990, Dar \& deRujula 2000) living in the cluster is also expected, as well as from
`active' galaxies (Urry \& Padovani 1995) which belong to the cluster.

In this paper, we report the results of a detailed spatial and spectral analysis of the unidentified EGRET sources
at high galactic latitude and the findings of a preliminary evidence for a correlation between galaxy clusters and
unidentified EGRET sources at $|b| > 20$ $\deg$. The plan of the paper is the following. In Sect.2 we discuss the
evidence for the spatial correlation between EGRET sources at high galactic latitude and galaxy clusters in the
Abell catalog (Abell et al. 1989). We also discuss the analysis of the gamma-ray flux and spectra of the EGRET
sources probably associated with galaxy clusters in comparison with those associated with other EGRET sources. In
Sect.3 we analyze in details each one of the 18 EGRET sources which have been found as possible candidates for
being physically associated with galaxy clusters. We finally derive a sample of 9 EGRET sources which are most
probably associated with 12 galaxy clusters. In Sect.4 we discuss the correlation we found between the gamma-ray
flux of the EGRET source and the radio flux of the cluster radio sources for the 9 most probable EGRET-cluster
associations and in Sect.5 we discuss the evidence for a similar correlation between the gamma-ray luminosity and
the X-ray luminosity of the same most probable EGRET-cluster associations. We present in Sect.6 our conclusions
and a discussion of the future expectations for the detection of gamma-ray emission from galaxy clusters in the
light of the next generation space and ground-based gamma-ray experiments. We use $H_0 = 50 ~{\rm km~ s}^{-1}
~{\rm Mpc}^{-1}$ and a flat ($\Omega_0 = 1$) cosmology throughout the paper unless otherwise specified.

\section{The spatial correlation of galaxy clusters with unidentified EGRET gamma-ray sources}

Motivated by the previous arguments, we analyzed the available data for the gamma-ray sources in the Third EGRET
catalog (Hartman et al. 1999) and we looked for a correlation between the position of unidentified gamma-ray
sources with $|b|>20$ $\deg$ and the positions of galaxy clusters in the Abell catalogue (Abell et al. 1989). We
further looked for the X-ray information about the selected clusters in the ROSAT all sky survey and pointed
observations and in the BeppoSAX cluster catalogue. We also looked for radio sources associated with galaxy
clusters in the NVSS radio survey, in the VLA surveys as well as in the available literature.\\
 We first studied such a spatial correlation within a fixed radius (1 $\deg$) from the center of each EGRET source
 and subsequently we refined our analysis considering the actual $95 \%$ confidence level position error contours of each EGRET
 source found in the previous step.

We found that 50 EGRET sources at high galactic latitude, $|b|>20$ $\deg$, are spatially correlated - within 1
degree from the center of the EGRET source - with the position of 70 galaxy clusters in the Abell catalogue (Abell
et al. 1989). We choose a correlation radius of 1 $\deg$ because this is the angular distance at which EGRET
cannot distinguish two separate point-like sources (Hartman et al. 1999). We performed a Monte Carlo simulation to
check if such a spatial association can be understood as a simple random projection effect. Specifically, we built
$10^3$ random distributions of galaxy clusters extracted from the Abell catalogue and we cross-correlated their
positions with the EGRET source positions within 1 $\deg$ radius.
We find that, on average, $33$ EGRET sources can be randomly associated with simulated cluster positions. Based on
a Kolmogorov-Smirnov test, the probability that all of the remaining $17$ EGRET unidentified sources are still
randomly associated with galaxy clusters is $\simlt 0.5 \%$.
 This indicates that the confidence level of the spatial association is about $2.96 \sigma$ (assuming a Gaussian
statistics which is justified for $\simgt 20$ spatial associations).

For a more detailed analysis, we correlated the positions of the Abell clusters with the exact position error
contours given for each EGRET source found in the Third EGRET catalogue. In this procedure we consider also the
spatial extension of the galaxy clusters. We find that the coordinates of the optical centers of 52 Abell clusters
fall within the contour containing the $95 \%$ confidence level error region for the positions of $39$ EGRET
sources. In this analysis we consider a positive correlation also for those clusters whose optical centers are
close to the border of the $95 \%$ confidence level error contours of the EGRET sources and whose spatial
extension is found within the $95 \%$ confidence level EGRET position error contour.
We then simulated, as before, $10^3$ random distributions of galaxy clusters extracted from the Abell catalogue
and we cross-correlated their positions with the EGRET source positions within their $95 \%$ confidence level
contours, finding that, on average, $26$ EGRET sources can be randomly associated with simulated cluster
positions. Based on a Kolmogorov-Smirnov test, the probability that all of the remaining $13$ EGRET unidentified
sources are still randomly associated with galaxy clusters is $\simlt 1 \%$ (or, in other words, the significance
of the probable correlation between galaxy clusters and EGRET sources is at more than $2.5 \sigma$ confidence
level).

Since a substantial fraction of the sky observed by EGRET has a low sensitivity (where it would be difficult to
observe any faint source), the previous estimate of the significance level of the correlation can be safely
considered as a lower limit of the true one.
 In fact, since the Abell galaxy cluster distribution is approximately uniform on the
 sky, the cluster -- EGRET source correlation we found here is suffering from a lack of other possible EGRET-cluster
 associations coming from those gamma-ray sources which are not detected in the low-exposure region of the EGRET sky.
 Assuming that the number of additional EGRET sources detectable with a uniform sky coverage, $N_{\rm x} \propto A_{\rm low-exp.}$
 (where $ A_{\rm low-exp.}$ is the area of the gamma-ray sky with low-exposure), is correlated with galaxy clusters in the same
 ratio of our previous estimates, and assuming that the fraction of random correlation is again similar to what
 previously estimated (i.e. $\sim 2/3$ of the correlations are random and $\sim 1/3$ are probable), the statistical
 confidence level of the correlation found after correcting for the non-uniform exposure of EGRET increases with
 increasing value of $N_{\rm x}$ and scales like $\sim \sqrt{N_{\rm x}} \propto \sqrt{A_{\rm low-exp.}}$, for high values of $N_{\rm x}$.
So, in conclusion, we believe that the previous estimate of the statistical significance of the cluster--EGRET
source correlation given above can be reliably considered as a lower limit to the actual significance level of the
spatial correlation between galaxy clusters and unidentified EGRET sources.

To select out of the full list previously found the more probable associations of galaxy clusters with the
unidentified EGRET sources, we superposed the optical cluster positions and their X-ray images onto the maps
containing the probability distribution for the spatial position of the 50 EGRET sources found in our spatial
correlation analysis.\\
We found that 18 of the original 50 EGRET sources associated with galaxy clusters have also an AGN (with confirmed
identification) whose position falls within the $95 \%$ confidence level position error contours of the gamma-ray
source. We also found that a Gamma Ray Burst is found in association with the EGRET source 3EG J2255-5012 and the
clusters A1073 -- A1074. Also a  SN remnant is found in the field of the source 3EG J1235+0233 associated to the
cluster A1564. We then excluded these 20 EGRET sources and the associated 30 clusters from the list of probable
physical associations.\\
 We also excluded 12 EGRET sources with a possible, but not confirmed, AGN contamination
in the Third EGRET catalog (see Hartman et al. 1999).
Note that also this procedure is very conservative since there are 4 cases out of the 12 listed in which the
possible AGN source is found beyond the $95 \%$ confidence level position error contours of the associated EGRET
sources, while the galaxy clusters spatially associated with the EGRET sources fall within their $95 \%$
confidence level position error contours.

Finally, we found in our conservative analysis that 24 galaxy clusters are associated to 18 unidentified EGRET
sources with $|b|>20$ $\deg$ for which there is no firmly established counterpart at other wavelengths, neither
extragalactic (AGN or ``active'' galaxy) or galactic (Supernova remnant, pulsar, neutron star). All of these
galaxy clusters have their optical and X-ray centers falling within the $95 \%$ confidence level position error
contours of the EGRET sources. We show in Table 1 the list of the 18 EGRET sources and the 24 clusters which are
spatially correlated within the $95 \%$ confidence level position error contours of each EGRET source. This is the
initial sample of likely associations between galaxy clusters and EGRET gamma-ray sources on which we performed a
more detailed analysis, as discussed in the following.

According to our selection procedure, the significance level of such a spatial association is $\approx 2.55
\sigma$ which corresponds to a probability $\simlt 1 \%$ for the null hypothesis that the two source populations
are randomly associated. However, the point is still to determine how many of these spatial associations are due
to random projection effects and which are the most probable physical associations. A rough estimate of the
probability to have still random associations in the sample here selected (see Table 1) and to be not contaminated
by either extra-galactic (AGN, GRB) or galactic (SNR, pulsars, ..) gamma-ray sources, yields that about $2/3$ of
the 18 selected EGRET sources are still random associations. This rough estimate would yield 6 most probable
cluster--EGRET source associations with a confidence level of $1.73 \sigma$. Note, however, that this is again a
lower limit to the true statistical confidence of the correlation since the effect of the non-uniform EGRET sky
coverage has to be taken into account and would tend to increase the statistical significance level of the most
probable association. If we correct for the number of correlations expected in the fraction of the EGRET sky
($\sim 30 \%$ of the full sky) which has a flux limit below $F(>100 ~{\rm MeV}) \leq 6 \cdot 10^{-8} cm^{-2}
s^{-1}$, we obtain that the expected confidence level of the most probable associations raises from $1.73 \sigma$
to $2.12 \sigma$.

\subsection{Flux and spectral analysis}

In addition to the spatial information contained in the Third EGRET catalog and in the Abell cluster survey, we
can use more physical criteria to determine the number of spurious correlations in our selected sample of Table 1.
Specifically, we first analyze the flux level, the flux variability and the spectral indices of the 18 EGRET
sources in Table 1 compared to the same quantities of other gamma-ray sources more definitely identified in the
Third EGRET catalogue (mainly AGN and Pulsars). Then we run Monte Carlo simulations of flux level and variability
for the probable EGRET--cluster associations to determine the fraction of random correlations expected in our
selected sample.
\begin{figure}[tbh]
\begin{center}
\hbox{
 \epsfig{figure=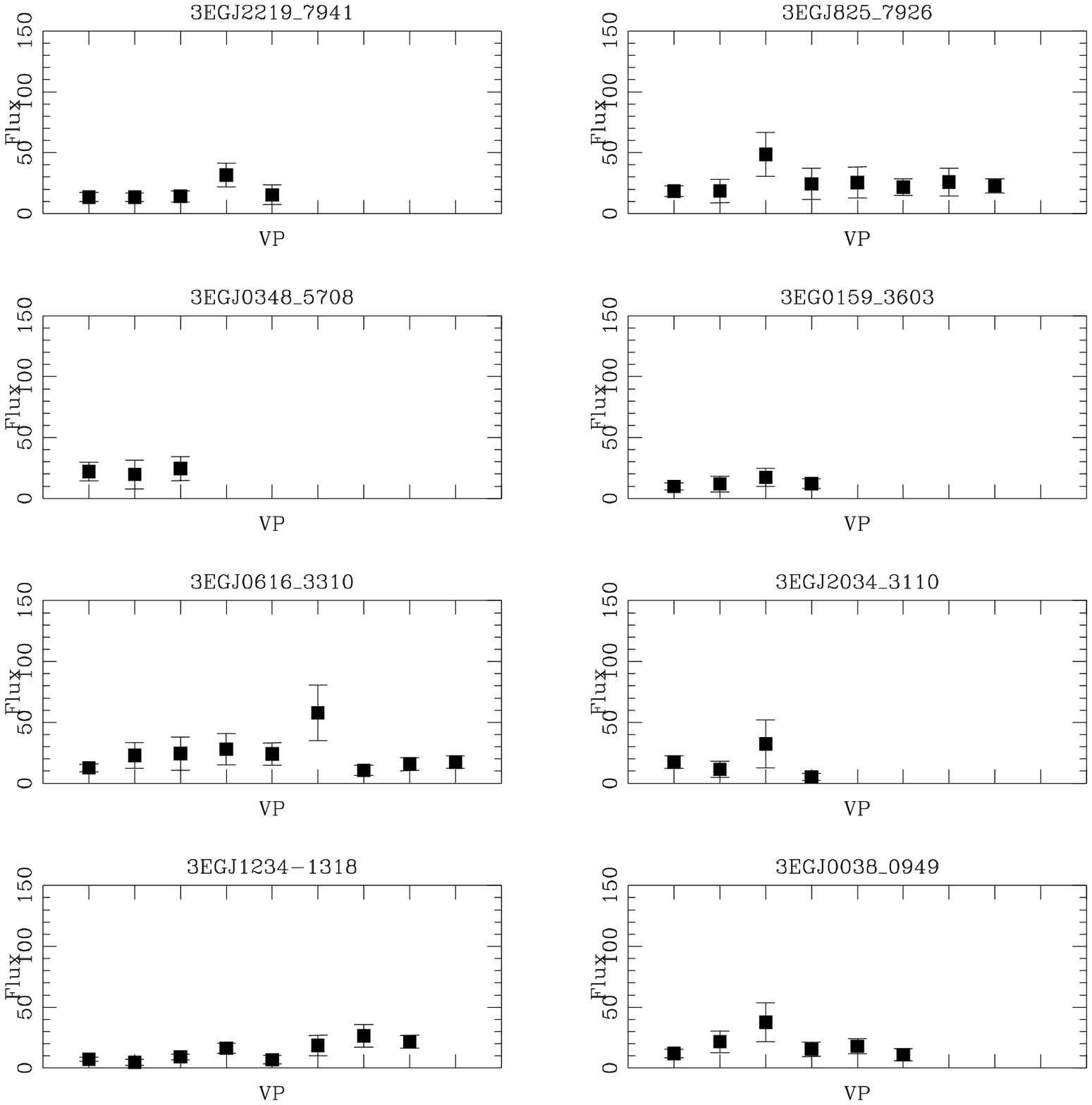,width=7.5cm,height=12.cm,angle=0.}
 \epsfig{figure=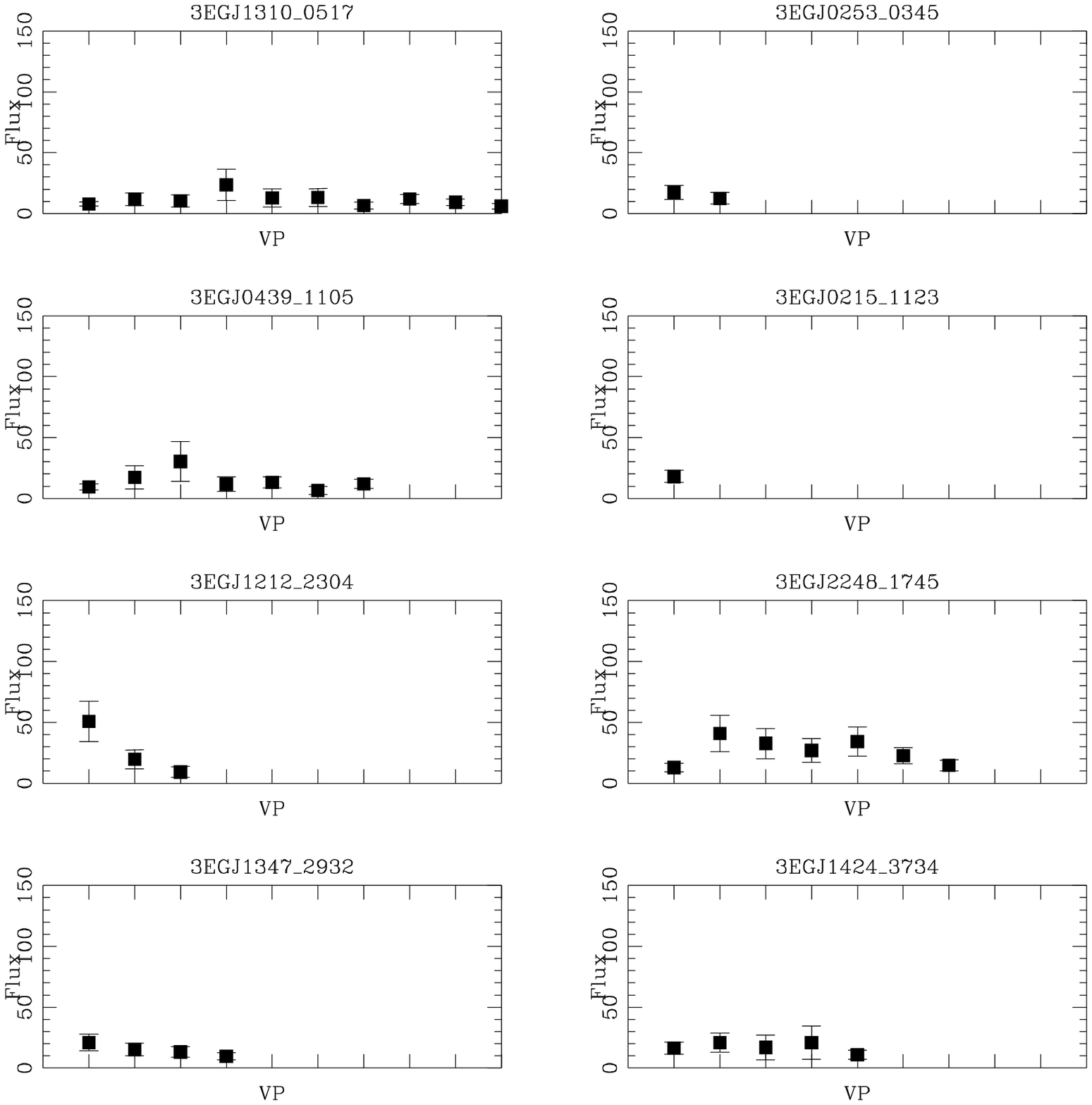,width=7.5cm,height=12.cm,angle=0.}
 \epsfig{figure=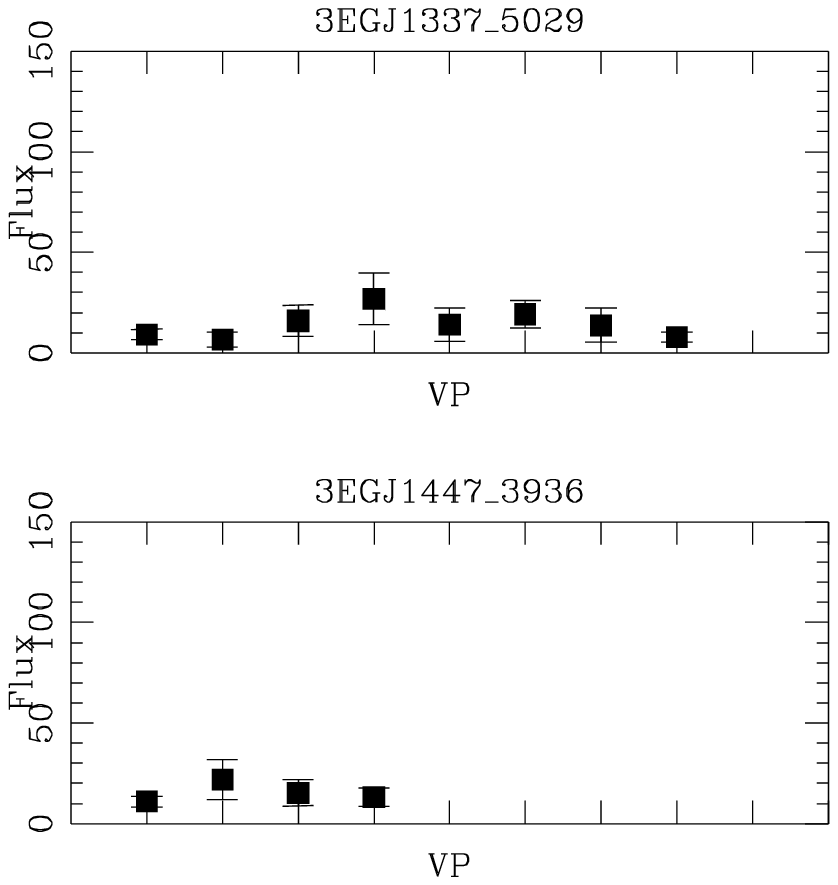,width=7.5cm,height=12.cm,angle=0.}
 }
\end{center}
\caption{\footnotesize {We show the gamma-ray flux of the EGRET sources in Table 1 as detected in the different
viewing periods (VP) of the source detection. Data are from Hartman et al. (1999). The flux detected in the
different VPs are reported here in the sequential order given in the Third EGRET catalog, being the correct
observing time sequence irrelevant for our purposes. The fluxes of the EGRET sources are in units of $10^{-8}$
cm$^{-2} s^{-1}$.
}}
 \label{figure: 1}
\end{figure}

Fig.1 shows the flux variation in the viewing periods (hereafter VP) over which the EGRET sources reported in
Table 1 have been detected. We notice that the flux variability for the probable cluster-EGRET source associations
listed in Table 1 is, on average, $\simlt 20 \%$ and only in a few cases (3EG J1825-7926, 3EG J1212+2304, 3EG
J0616-3310, 3EG J2248+1745) it can be considered $\simgt 30 \%$ in some specific VP (see Fig.1). The
correspondingly associated clusters (see Table 1) are poorly studied, do not have X-ray information and do not
have any identified bright radio galaxy or radio halo/relic emission. Hence, we also consider these cases as
suspiciously due to projection effects.
 Beyond the positive detections with high statistical significance $(TS)^{1/2} \simgt 4$ of the EGRET sources
 reported in Fig.1, the Third EGRET catalog provides also upper limits on their fluxes in other independent VPs.
 Such upper limits have $(TS)^{1/2} < 2$ (i.e., a low statistical significance) and we verified
 that most of them are consistent with the positive detections of the EGRET sources we show in Fig.1.
 In some cases, however, (see, e.g., 3EGJ0348-5708, 3EGJ1234-1318, 3EGJ0253-0345, 3EGJ0215+1123)
 there are upper limits which are well below the flux level found in other independent VP
 detections of the sources. Nonetheless, we noticed that these ``quite low'' upper limits have all a very low
 statistical confidence level, $(TS)^{1/2} \sim 0$, and are hence extremely unreliable.
 Thus, due to their quite low statistical significance, the upper limits of the EGRET sources listed in Table 1 and
 shown in Fig.1 do not strongly affect our conclusions on their overall flux variability.
 A few other sources with independent flux upper limits below the definite detections (see, e.g., 3EGJ0616-3310,
 3EGJ2034-3110, 3EGJ1212+2304) show also a level of flux variability which does not justify to consider them
 as stationary sources.
For the sake of completeness, we will discuss in Sect.3 below the detailed analysis of each specific EGRET source
listed in Table 1.\\
 For comparison, we show in Fig.2 the flux variation of the EGRET sources which are correlated with galaxy
 clusters and moreover contain also a confirmed AGN in the field.
 In these last cases, the flux of the EGRET sources not only show stronger and statistically significative variations,
 but also have a much higher value of their average gamma-ray flux.
\begin{figure}[tbh]
\begin{center}
\hbox{
 \epsfig{figure=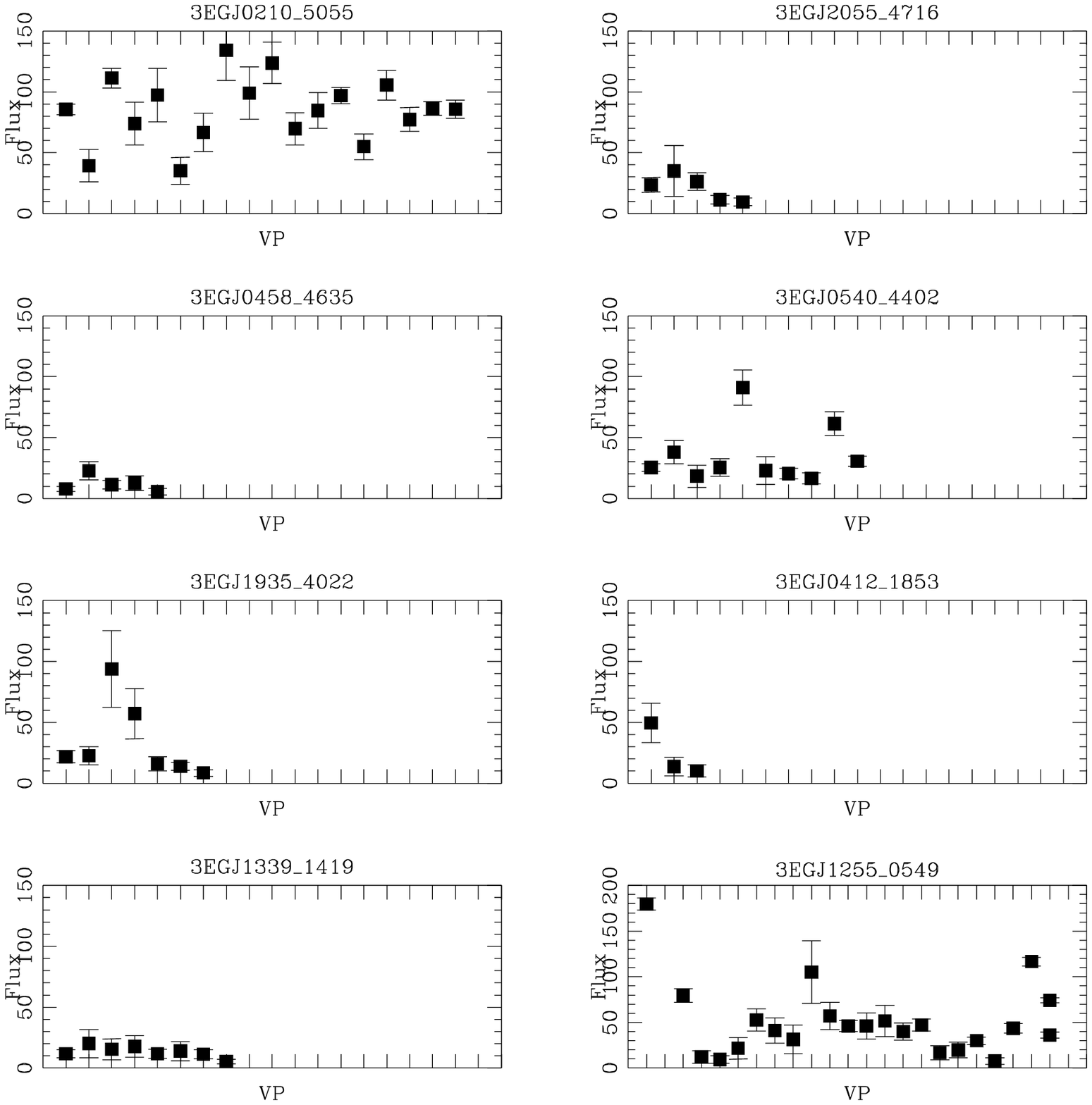,width=7.5cm,height=12.cm,angle=0.}
 \epsfig{figure=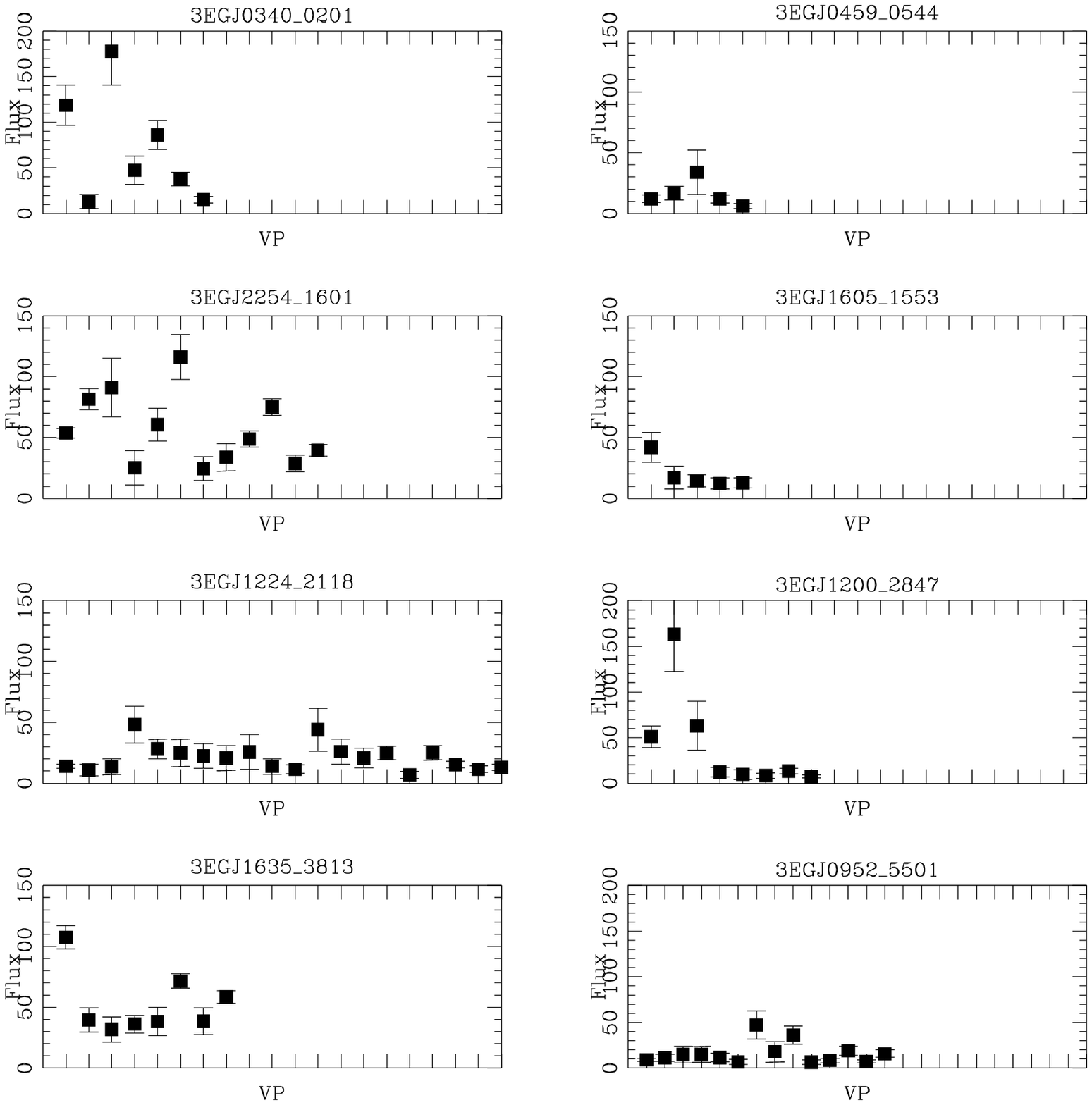,width=7.5cm,height=12.cm,angle=0.}
 \epsfig{figure=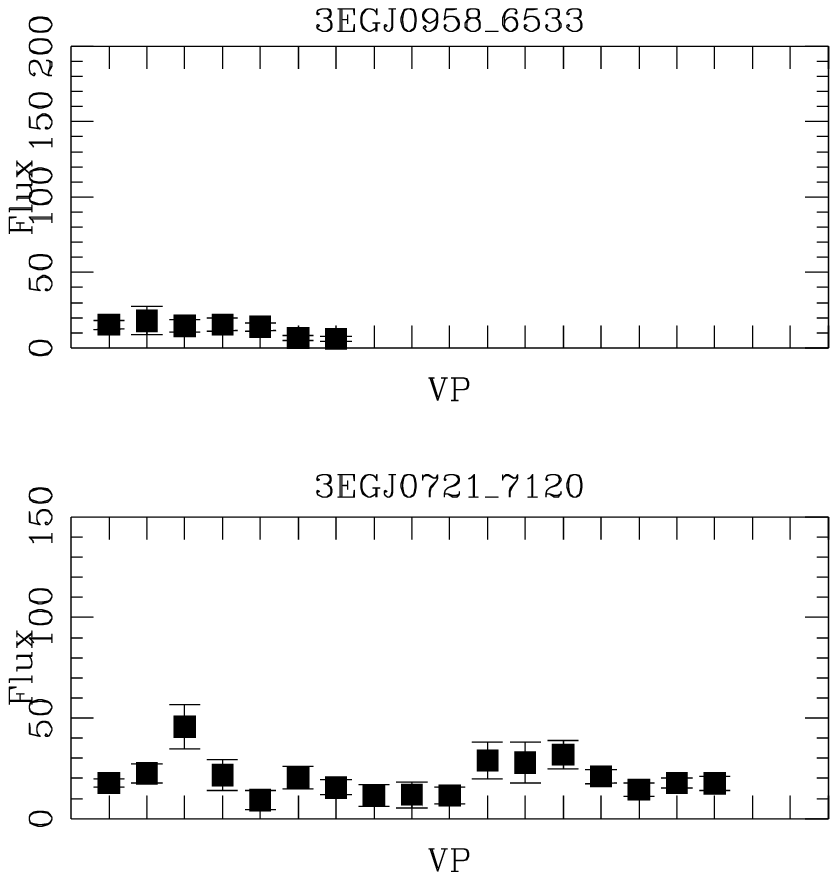,width=7.5cm,height=12.cm,angle=0.}
 }
\end{center}
\caption{\footnotesize {The variation of the gamma-ray flux of the EGRET sources which are correlated with galaxy
clusters and which have also an identified AGN in the field. Data are from Hartman et al. (1999). As in Fig.1, the
flux detected in the different VPs are reported in the sequential order given in the Third EGRET catalog, being
the correct observing time sequence irrelevant for our purposes. The fluxes of the EGRET sources are in units of
$10^{-8}$ cm$^{-2} s^{-1}$.
}}
 \label{figure: 2}
\end{figure}

In Fig.3 we compare the spectral index, $\gamma$, of the EGRET sources which are probably associated with galaxy
clusters with those of the EGRET sources which are spatially correlated with galaxy clusters and moreover contain
an AGN in the field. The EGRET sources correlated with clusters are not found to be brighter than $F(>100 {\rm
MeV}) \sim 2 \cdot 10^{-7}$ counts cm$^{-2} s^{-1}$ and show spectral indices in a large range $\sim 2 - 3.5$.
With a remarkable difference, the EGRET sources identified with known AGNs span over a much higher gamma-ray flux
range and have a much smaller range of spectral index values ($\gamma \sim 2 - 2.5$) especially at very bright
flux levels $F(>100 ~{\rm MeV}) >  5 \cdot 10^{-7}$ counts cm$^{-2} s^{-1}$. Pulsars also show very flat spectral
indices $\gamma \simlt 2$ and very high gamma-ray flux which cannot be compared with those of the EGRET sources
associated with clusters.

The gamma-ray spectral indices for the probable associations listed in Table 1 have values which are consistent
with those expected from the viable mechanisms for gamma-ray emission in clusters. Theoretical models for cluster
gamma-ray emission predict in fact slopes in the range $\gamma \sim 1.8 - 3.2$, going from annihilation of dark
matter neutralinos (Colafrancesco \& Mele 2001) to non-thermal electron bremsstrahlung (Colafrancesco 2001a,b,
Blasi 2000). Only the sources 3EG J2034-3110 (associated to A886) and 3EG J1424+3734 (associated to A1902-A1914)
have spectral indices $\simgt 3$, even though with large uncertainties. However, while the first source, 3EG
J2034-3110, shows also some level of flux variability (see Fig.1) and could then be contaminated by AGN-like
sources, the gamma-ray source 3EG J1424+3734 has a very low flux variability ($\sim 15 \%$) and is likely to be a
probable association whose gamma-ray emission could be dominated by non-thermal electron bremsstrahlung, which
shows typically a steep spectrum consistent with that of the parent cosmic-ray electrons (see, e.g., Longair
1993).
\begin{figure}[tbh]
\begin{center}
\epsfig{figure=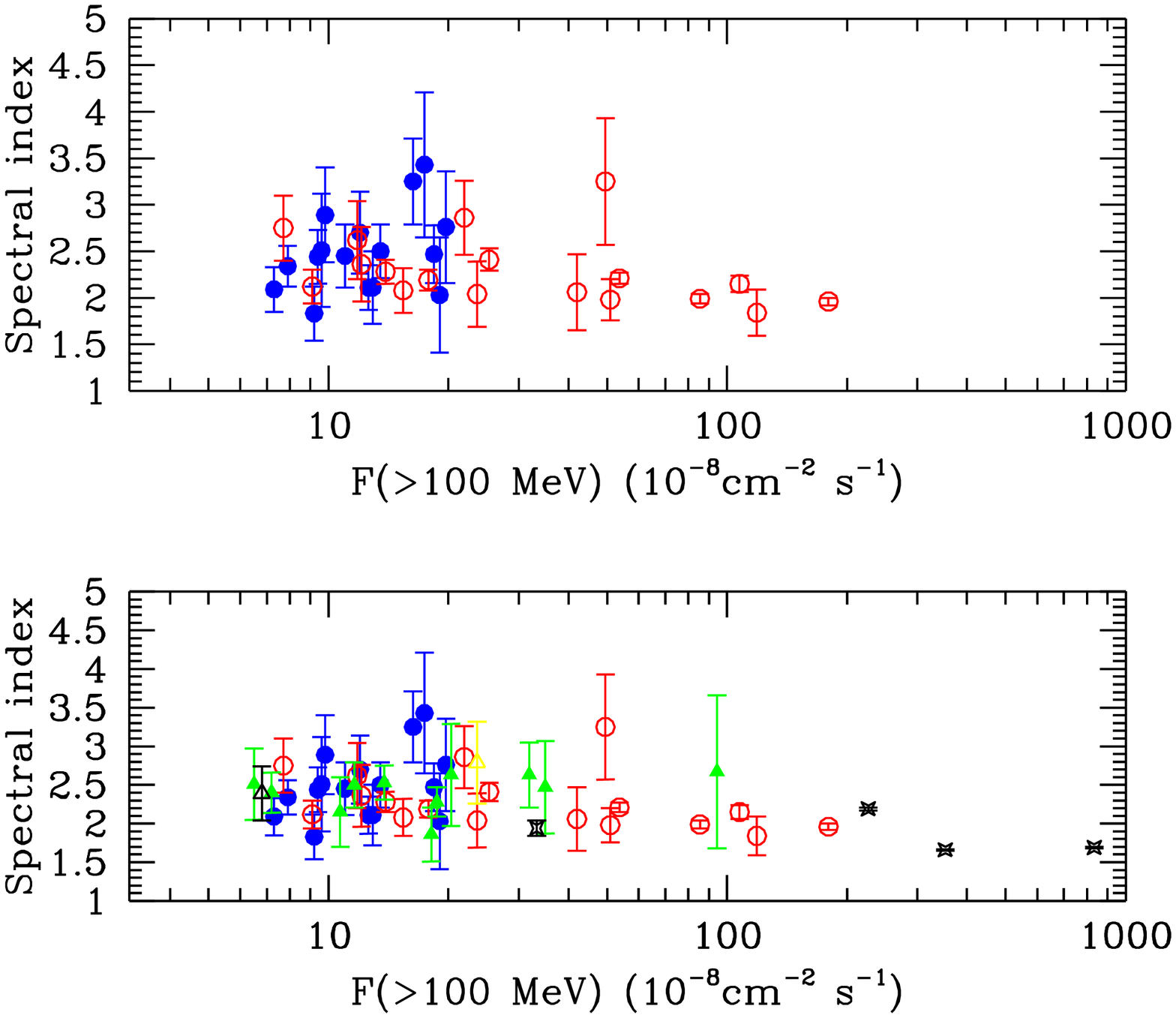,width=12.cm,angle=0.}
\end{center}
\caption{\footnotesize {The gamma-ray spectral index $\gamma$ is plotted against the gamma-ray flux $F(>100 {\rm
MeV})$ of the EGRET sources found to be spatially correlated with galaxy clusters. In the top panel we compare the
spectral indices for the EGRET sources which are more probably associated with clusters (filled circles) with the
EGRET sources again correlated with clusters  and in which an identified AGN is also found (open circles). In the
lower panel the data from the EGRET sources contaminated by possible AGNs (gray triangles), GRB (open, light-gray
triangle), SNR (open black triangle) and Pulsars (stars) are added to the previous data sets. Data are from
Hartman et al. (1999).}}
 \label{figure: 3}
\end{figure}

We finally run Monte Carlo simulations of the flux variability level of the 18 EGRET sources of Table 1. For a
uniform random distribution of their fractional flux change, $\Delta F/F$, we expect 4 EGRET sources with $\Delta
F/F \simlt 0.2$, while the remaining 14 EGRET sources possibly associated with galaxy clusters should have $0.2
\simlt \Delta F/F \simlt 1$.  The actual data reported in Fig.1 show that there are about 11 EGRET sources with
$\Delta F/F \leq 0.2$ and only 7 sources with $0.2 \simlt \Delta F/F \simlt 1$. This indicates that the low flux
variability shown by the EGRET sources found in association with clusters cannot be recovered by a simple random
distribution at more than $5 \sigma$ confidence level.

Based on these results we expect that about 10 EGRET sources out of the 18 listed in Table 1 are probable
EGRET-cluster associations  having $\Delta F/F \simlt 0.2$, $F(>100 ~{\rm MeV}) < (1 - 2)\cdot 10^{-7} cm^{-2}
s^{-1}$ and $\gamma \sim 2 - 3.2$. However, only a detailed analysis of the spatial and spectral features of each
EGRET source as well as a detailed analysis of their cluster counterparts can reveal the nature of the more
probable physical association. We will present in the next section the detailed analysis of each one of the
specific EGRET sources listed in Table 1 and of their possible astrophysical counterparts.

\section{Analysis of the specific sources}

In this section we analyze in details the more probable associations of galaxy clusters with the unidentified
EGRET sources which are listed in Table 1: all of these galaxy clusters are found within the $95 \%$ confidence
level position error contours of the associated EGRET source. As for the flux of each source, we report only the
first entry given in the Third EGRET catalogue (Hartman et al. 1999) providing also the viewing period (VP) of the
source detection and the significance level, $(TS)^{1/2}$, of the detection. The reader may refer to Hartman et
al. (1999) for the full list of information about the EGRET source under consideration.
\begin{figure}
\begin{center}
\hbox{ \vbox{\hbox{
  \parbox{8.5cm}{\fbox{a85-bw-last.GIF}}
  \parbox{8.5cm}{\fbox{a388-bw-last.GIF}}
   }
\vskip 8.5truecm
   \hbox{
     \parbox{8.5cm}{\fbox{a1914-bw-last.GIF}}
      \parbox{8.5cm}{\fbox{a1024-bw-last.GIF}}
      }
\vskip 8.5truecm
      \hbox{
       \parbox{8.5cm}{\fbox{a1688-bw-last.GIF}}
      }
} }
\end{center}
\caption{\footnotesize {We show here the positions of the galaxy clusters which are more probable candidates for
the associations with unidentified EGRET sources: 3EGJ0038-0949 associated with A85 (upper left), 3EGJ0253-0345
associated with A388 (upper right), 3EGJ1424+3734 associated with A1914-A1902 (mid left), 3EGJ2218-7941 associated
with A1024-A1014 (mid right) and 3EGJ1310-0517 associated with A1688 (bottom left).
 The cluster positions and X-ray brightness contours (when available) are superposed to the maps of the EGRET sources.
 The intensity scale of the EGRET maps goes from black (minimum) to white (maximum). }} \label{figure: 4}
\end{figure}

\subsection{3EG J2219-7941}

This EGRET source has been detected with a $(TS)^{1/2} = 4.4$ in the VP=P1234. It has a flux $F_{P1234}(>100 {\rm
MeV}) =(13.5 \pm 3.6) \cdot 10^{-8} ~{\rm cm}^{-2}~{\rm s}^{-1}$ with a power-law spectral index $\gamma=2.50 \pm
0.29$. Its flux does not change significantly over all the VPs with the exception of the VP=10.0 in which a flux
increase by a factor $\sim 2$ has been recorded; however, this flux is consistent with the flux detected in the
other periods at the $2 \sigma$ confidence level, so that it can be considered a stationary gamma-ray source (see
Fig.1). There is no identified gamma-ray source counterpart for this EGRET source and, thus, it is a good
candidate for the galaxy cluster association. Two Abell clusters fall within the $95 \%$ confidence level position
error contours of this source: A1014 and A1024. The EGRET source map is very broad with an effective radius of the
$95 \%$ confidence level position error circle of $\theta_{95} = 0.63$ $\deg$. The elongation of the EGRET source
probability map is aligned with the position of the two clusters, suggesting a possible contribution to the
gamma-ray emission from both clusters (see Fig.4).\\
 Both these clusters have X-ray and radio information.
 A1014 at $z=0.1165$ (Struble et al. 1999) has been found in the
RASS with a flux $F_{0.5-2.5 {\rm keV}} \simlt 6.97 \cdot 10^{-13} ~{\rm erg} {\rm cm}^{-2} ~{\rm s}^{-1} $ (Briel
et al. 1993). The cluster A1014 has also several NVSS radio sources located in its vicinities.\\
 A1024 has an optical redshift of $z=0.0734$ (Struble et al. 1999) and is associated with the ROSAT X-ray source RX J1028.3+0345.
 Such X-ray source has a flux $F_{0.1-2.4 ~{\rm keV}} =
(0.158 \pm 0.040) \cdot 10^{-11} ~{\rm erg} ~{\rm cm}^{-2} ~{\rm s}^{-1}$ in the $0.1 - 2.4$ keV band (Brinkman et
al. 1995). These last authors also reported a radio flux at $5$ GHz, $F_{5 ~{\rm GHz}} = 0.127 \pm 0.019$ Jy, as
taken from the Gregory \& Condon (1991) catalogue. The cluster A1024 is physically associated to three radio
galaxies ({\it PMN J1028+0345}, {\it MG1 J102825+0345}, {\it RGB J1028+037}) found at the same redshift of the
galaxy cluster and there are also several other NVSS radio sources in the field. One of the radio galaxies, namely
{\it 1025+040} found in the White \& Becker (1992) catalog at $z=0.0733$, is a wide angle tail (WAT) radio galaxy
(Sakelliou \& Merrifield 2000). A1024 has been also observed with the VLA in the B configuration and a radio flux
at $1.4$ GHz, $S_{1.4}= 272$ mJy with a radio power $P_{1.4}=24.48$ ~W ~Hz$^{-1}$ (probably due to the WAT radio
galaxy) has been detected (Owen \& Ledlow 1997).\\
 Based on the previous evidence, we consider that this is a probable association between galaxy clusters and
 an EGRET gamma-ray source.

\subsection{3EG J1825-7926}

This EGRET source has been detected with a $(TS)^{1/2} = 4.9$ in the VP=P1234. It has a flux $F_{P1234}(>100 ~{\rm
MeV}) =(18.4 \pm 4.5) \cdot 10^{-8} ~{\rm cm}^{-2} ~{\rm s}^{-1}$ with a power-law spectral index $\gamma=2.47 \pm
0.31$. Its flux does not change significantly over all the VPs with the exception of the VP=38.0 in which a flux
increase by a factor $\sim 2$ has been recorded; however, this flux is consistent with the flux detected in the
other periods at less than the $2 \sigma$ confidence level (see Fig.1). The quite low upper limit of $< 13.5 \cdot
10^{-8} ~{\rm cm}^{-2} ~{\rm s}^{-1}$ found in the VP=402.+  has $(TS)^{1/2} = 0$ and thus is not statistically
significant.
 The Abell cluster A3631 falls at the border of the $95
\%$ confidence level position error contours of the EGRET source. The position error map of this EGRET source is
quite broad with $\theta_{95} = 0.78$ $\deg$ and it is elongated in the south-west north-east direction. There is
no other known gamma-ray source counterpart for this EGRET source. However, there is poor information available on
the cluster A3631. In particular, there is no radio source found in the field of this cluster and no X-ray
information.\\
 Due to the previous evidence, there is no strong hint indicating the possible association of this cluster with the EGRET source 3EG
J1825-7926 and so we consider this case as likely due to projection effects.

\subsection{3EG J0348-5708}

This EGRET source has been detected with a $(TS)^{1/2} = 4.1$ in the VP=P2. It has a flux $F_{P2}(>100 ~{\rm
MeV})=(22.1 \pm 7.6)\cdot 10^{-8} ~{\rm cm}^{-2} ~{\rm s}^{-1}$ but the power-law spectral index remains
unconstrained. Its flux does not change over all the VPs in which it has been detected and it is a stationary
gamma-ray source (see Fig.1).
 The quite low upper limits $ <10 \cdot 10^{-8} ~{\rm cm}^{-2} ~{\rm s}^{-1}$ found in other independent VPs
have all $(TS)^{1/2} = 0$ and can not be considered as statistically significant. The other reported upper limits
on this source are consistent with the detection fluxes.\\
 The Abell cluster A3164 falls within the $95 \%$ confidence level position
error contours of the EGRET source.
 Also, there is no other identified gamma-ray source counterpart for this EGRET source.
The EGRET source map is relatively broad with $\theta_{95} = 0.42$ $\deg$ and has a comet-like tail in the east
direction.

A3164 is an irregular cluster with a redshift of $z = 0.057$ (Struble et al. 1999) for which there is poor
information available. Ebeling et al. (1996) estimated its X-ray luminosity to be $L_{0.5-2.4~{\rm keV}} \approx
1.48 \cdot 10^{44}$ erg s$^{-1}$ and its temperature as $kT \approx 3.6$ keV from the ROSAT data. There are no
evidence for NVSS radio sources found in correlation with this cluster.\\
  Based on the previous evidence, we do not find any strong hint for the probable association of this
cluster with the EGRET source 3EG J1825-7926 and we consider also this case as likely due to projection effects.

\subsection{3EG J0159-3603}

This EGRET source has been detected with a $(TS)^{1/2} = 4.3$ in the VP=P1234. It has a flux $F_{P1234}(>100 {\rm
~MeV})=(9.8 \pm 2.8)\cdot 10^{-8} ~{\rm cm}^{-2} ~{\rm s}^{-1}$ with a power-law spectral index $\gamma=2.89 \pm
0.51$. Its flux does not change over all the VPs and it is a stationary gamma-ray source (see Fig.1).
 Two Abell clusters fall within the $95 \%$ confidence level position error contours of this source:
 A219 and A2963. Other galaxy clusters are found in the vicinities of the EGRET source (see Fig.5).
 The EGRET
source map is quite broad with $\theta_{95} = 0.79$ $\deg$ and the elongation of the EGRET probability map is
aligned with the position of the two clusters, suggesting a possible contribution to the gamma-ray emission from
both clusters. There is no other known gamma-ray source counterpart for this EGRET source.

There is little optical and X-ray information on both the clusters A219 and A2963. There are nonetheless three
NVSS radio sources associated with the cluster A2963: they have radio flux at $1.4$ GHz of $S_{1.4}=
(23.44\pm1.57), (4.46 \pm 0.45)$ and $(2.43\pm0.45)$ mJy, respectively.\\ In view of these evidence, we consider
that this is a candidate for a probable association between galaxy
 clusters and an EGRET gamma-ray source.

\subsection{3EG J0616-3310}

This EGRET source has been detected with a $(TS)^{1/2} = 4.7$ in the VP=P1234. It has a flux $F_{P1234}(>100 {\rm
~MeV}) =(12.6 \pm 3.2) \cdot 10^{-8} ~\cm2s$ with a power-law spectral index $\gamma=2.11 \pm 0.24$. Its flux
changes significantly over several VPs and in the VP=419.5 it increase by a factor $\simgt 4$ with respect to the
VP=P1234. Due to such strong flux variations in comparison with other cases shown in Fig.1, it is hard to consider
it as a stationary gamma-ray source.

Two clusters (A577 and A575) fall close to the $95 \%$ confidence level error contour for the position of this
EGRET source. Another cluster (A573) falls within 1 $\deg$ radius from the center of the EGRET source. However,
the shape of the EGRET map of this source is quite compact and round with $\theta_{95}=0.13$ $\deg$.\\
 The two clusters A577 and A575 have very few morphological and physical information (see Table 1).
 A575 has an estimated X-ray flux of $F_{0.1-2.4 ~{\rm keV}} = (1.9
\pm 16.8) \cdot 10^{-12} ~{\rm erg} ~{\rm cm}^{-2} ~{\rm s}^{-1}$ with an estimated luminosity of $L_{0.1-2.4
~{\rm keV}} \sim 9.3 \cdot 10^{43} ~{\rm erg} ~{\rm s}^{-1}$ (Boehringer et al. 2000).
  Nonetheless, the cluster A577 is associated
with three NVSS radio sources with flux $S_{1.4} = (4.32 \pm 0.48), (14.92 \pm 0.61)$ and $(12.48 \pm 1.57)$ mJy,
respectively. Also A575 is correlated with other three NVSS radio sources with flux $S_{1.4} = (5.86 \pm 0.48),
(14.17 \pm 0.59)$ and $(3.03 \pm 0.47)$ mJy, respectively. Note that also the cluster A573 is associated with 7
NVSS radio sources.\\
 Due to the previous evidence, and in particular the flux variation over the various VPs,
 we consider this association as suspect and probably due to projection effects.

\subsection{3EG J2034-3110}

This EGRET source has been detected with a $(TS)^{1/2} = 4.0$ in the VP=P1. It has a flux $F_{P1}(>100 ~{\rm
MeV})=(17.4 \pm 5.2) \cdot 10^{-8} ~\cm2s$ with a power-law spectral index $\gamma=3.43 \pm 0.78$. Even though the
flux variations over the different VPs and the upper limit of $<6.2 \cdot 10^{-8} ~\cm2s$ with $(TS)^{1/2} =0$
found in the VP=209.0  are not statistically significant, the behaviour of this EGRET source is quite different
from the other sources here selected as possible association with galaxy clusters, which are expected to be quite
stationary over different VPs. Due to such flux variations in comparison with other cases shown in Fig.1, we do
not consider it as a stationary gamma-ray source.

Nonetheless, this EGRET source is quite broad and irregular with a quite large value of $\theta_{95}= 0.73$
$\deg$. The cluster A886 falls within the $95 \%$ confidence level position error contour of the source. There is
no other gamma-ray source counterpart in the field of 3EG J2034-3110.\\
 The cluster A886 has no detailed information available, it is not associated with any
 NVSS radio sources and there are no other hints for the presence of galaxy activity in its environment.\\
 Due to the previous evidence, and in particular the flux variation over the various VPs,
 we consider also this association as suspect and probably due to projection effects.

\subsection{3EG J1234-1318}

This EGRET source has been detected with a $(TS)^{1/2} = 4.8$ in the VP=P1234. It has a flux $F_{P1234}(>100 ~{\rm
MeV})=(7.3 \pm 1.7) \cdot 10^{-8} ~\cm2s$ with a quite low power-law spectral index $\gamma=2.09 \pm 0.24$. The
flux variations over the different VPs are not statistically significant, and for this reason it can be considered
as a stationary gamma-ray source. The upper limit of $< 8.9 \cdot 10^{-8} ~\cm2s$ has $(TS)^{1/2} = 0$ and its
very poor statistical significance does not affect strongly the previous conclusion.

Two galaxy clusters (A1558 and A1555) fall within the $95 \%$ confidence level position error contour of the EGRET
source. This EGRET source is quite regular with $\theta_{95}= 0.76$ $\deg$ even though source confusion may affect
its flux and/or its position (see Hartman et al. 1999). No other known gamma-ray source counterpart has been found
in the field of this EGRET source.

A1558 has an estimated redshift of $z=0.145$  and it is associated with two NVSS radio sources with flux $S_{1.4}
= 2.92 \pm 0.47$ and $7.73 \pm 1.24$ mJy, respectively.
 Also the  cluster A1555 is associated with two NVSS radio sources with flux $S_{1.4} = 7.14 \pm 0.48$ and $4.24 \pm 0.48$ mJy, respectively.
 No other information is available on these clusters at both optical and X-ray frequencies.\\
 Due to the previous evidence, we find that the association of this EGRET source with the two Abell clusters here
mentioned is still questionable.
\begin{figure}[tbh]
\begin{center}
\hbox{ \vbox{\hbox{
   \parbox{8.5cm}{\fbox{a331-bw.GIF}}
    \parbox{8.5cm}{\fbox{a497-bw.GIF}}
    }
\vskip 8.truecm
    \hbox{
     \parbox{8.5cm}{\fbox{a2963-bw.GIF}}
     \parbox{8.5cm}{\fbox{a1758-bw.GIF}}
      }
\vskip 4.5truecm
 } }
\end{center}
\caption{\footnotesize {We show here the positions of the galaxy clusters which are probable candidates for the
associations with unidentified EGRET sources: 3EGJ0215+1123 associated with A331 (upper left), 3EGJ0439+1105
associated with A497 (upper right), 3EGJ0159-3603 associated with A2963 and A219 (lower left) and 3EGJ1337+5029
associated with A1758 (lower right). The cluster positions and the X-ray brightness contours (when available) are
superposed to the maps of the EGRET sources. The intensity scale of the EGRET maps goes from black (minimum) to
white (maximum).}} \label{figure: 5}
\end{figure}

\subsection{3EG J0038-0949}

This EGRET source has been detected with a $(TS)^{1/2} = 4.1$ in the VP=P1234. It has a flux $F_{P1234}(>100 ~{\rm
MeV}) =(12.0 \pm 3.7) \cdot 10^{-8} ~\cm2s$ with a power-law spectral index $\gamma=2.70 \pm 0.44$. The flux
variations over the different VPs are at less than the $2 \sigma$ level and so are not strongly statistically
significant.
 Also the low upper limit of $<11.8 \cdot 10^{-8} ~\cm2s$ obtained in the VP=327.0 has $(TS)^{1/2} = 0$ and it
 is not statistically significant.
 However, the behaviour of the flux changes in the different viewing periods over which this EGRET source
has been detected is somewhat different from a purely stationary source as shown in Fig.1.
 This EGRET source is elongated
in the east-west direction and has a value $\theta_{95}= 0.59$ $\deg$.

The optical and X-ray center of the cluster A85 is found slightly beyond the $95 \%$ confidence level position
error contour of the EGRET source (see Fig.4). However, due to the its large extension ($\simgt 30$ arcmin radius)
a large part of this nearby ($z=0.056$) cluster falls within the $95 \%$ confidence level position error contour
of the EGRET source and hence can be considered to be spatially correlated with it. No other known gamma-ray
source counterpart is found in the field of this EGRET source.\\
 A85 is a bright X-ray cluster with a luminosity $L_{2-10 ~{\rm keV}}=(7.65 \pm 0.52)\cdot 10^{44} ~{\rm erg} ~{\rm s}^{-1}$ and a temperature
 of $kT = 6.2 \pm 0.4$ keV (Wu et al. 1999) and shows a strong activity in its ICM.\\
In fact, there are several bright radio galaxies within the cluster A85 and also several bright NVSS radio sources
correlated with the cluster as well as in the field of the relative EGRET source. A85 has been observed with the
VLA in the B and C configurations and a flux $S_{1.4} = 55$ mJy  has been reported by Owen \& Ledlow (1997). A85
contains also a diffuse, relic radio source found off-center with respect to the X-ray center of the cluster (see
Fig.6). Giovannini \& Feretti (2000) estimated that the diffuse radio halo flux at $1.4$ GHz is $S_{1.4} = 46$
mJy, consistently with the result of Owen \& Ledlow (1997), with a power $P_{1.4} = 6.1 \cdot 10^{23}$ W
Hz$^{-1}$.\\
 There are also evidence of an hard X-ray emission excess which is spatially correlated with the radio relic
source and is due probably to Inverse Compton Scattering (ICS) of the CMB photons with the relativistic electrons
of the radio relic (Bagchi et al. 1998, Lima-Neto et al. 2001). Such a non-thermal X-ray emission is spatially
correlated with the Very Steep Spectrum radio source {\it MRC 0038-096} (see Bagchi et al. 1998), without any
detected optical counterpart, which is $\sim 7$ arcmin south-west of the X-ray center of the cluster.\\
 The positive detection of both synchrotron radio and ICS X-ray emission from a common ensemble of relativistic
 electrons leads to an estimate of the average magnetic field, $B \approx 0.95 \pm 0.10 \mu$G, on the cluster
 scale. Further, the radiative flux and the estimated value of $B$ imply the presence of relativistic electrons
 (with radiative lifetime $\simgt 10^9$ yr) with Lorentz factor $\gamma_L \approx 700 - 1700$ (Bagchi et al. 1998).
Electrons with these energies can easily emit gamma-rays at $E_{\gamma} > 100 ~{\rm MeV}$ by bremsstrahlung  in
addition to the ICS emission tail which is present in the gamma-ray region probed by EGRET.\\
 Even though the cluster A85
is offset with respect to the center of the EGRET source map, there are good reasons to believe that it may
contribute substantially to the gamma-ray flux of the EGRET source 3EG J0038-0949 in addition to the possible
gamma-ray flux possibly produced by the active radio-galaxies which are living in the cluster environment.
\begin{figure}[tbh]
\begin{center}
  \parbox{15.cm}{\fbox{a85fig6.JPG}}
\vskip 15.truecm
\end{center}
\caption{\footnotesize {One of the most probable associations between galaxy clusters and EGRET unidentified
gamma-ray sources: A85. This cluster has a radio halo/relic inhabiting the cluster and a number of identified
radio galaxies in the ICM. Shown are the EGRET image with the cluster X-ray brightness contours (right), the
ROSAT-HRI X-ray image of the cluster (center) with the NVSS radio sources in the field (red circles) and the radio
halo/relic image obtained with the VLA at 327 MHz (left). }} \label{figure: 6}
\end{figure}

\subsection{3EG J1310-0517}

This EGRET source has been detected with a $(TS)^{1/2} = 5.0$ in the VP=P1234. It has a flux $F_{P1234}(>100 ~{\rm
MeV})=(7.9 \pm 1.8) \cdot 10^{-8} ~\cm2s$ with a power-law spectral index $\gamma=2.34 \pm 0.22$. The flux
variations over the different VPs are at less than the $2 \sigma$ level and are not statistically significative
(see Fig.1). Also the lowest upper limit $<10.9 \cdot 10^{-8} ~\cm2s$,
 obtained for this source in the VP=Virgo4 with $(TS)^{1/2} = 0.5$ has a very poor statistical significance.
  The EGRET source is elongated in the
south-north direction and has a value $\theta_{95}= 0.78$ $\deg$.

The center of the cluster A1688 is found within the $95 \%$ confidence level position error contour of the source,
even though quite off-center with respect to the EGRET map center (see Fig.4). There is, however, no other known
gamma-ray source counterpart in the field of this EGRET source.\\
 A1688 is one of the most distant clusters listed in Table 1 and has little information available at both optical
 and X-ray wavelengths.
 Kowalski et al. (1984) gave an estimate of its redshift $z\sim 0.19$ and of its X-ray luminosity, $L_{2-10 ~{\rm keV}}
\simlt 4.79 \cdot 10^{44}$ erg s$^{-1}$ as obtained from the HEAO-A1 all-sky survey. There are, nonetheless, 4
NVSS radio sources correlated with the position of A1688 and they have flux $S_{1.4} = 8.63 \pm 0.50; 59.83 \pm
2.25; 10.29 \pm 0.53; 3.66 \pm 0.51$ mJy, respectively.\\
 Due to these evidence, this could be considered as a probable
- but still questionable - association between galaxy clusters and EGRET gamma-ray source.

\subsection{3EG J0253-0345}

This EGRET source has been detected with a $(TS)^{1/2} = 4.0$ in the VP=317.0. It has a flux $F_{317.0}(>100 ~{\rm
MeV}) =(17.3 \pm 5.7) \cdot 10^{-8} ~\cm2s$ with a power-law spectral index which is unconstrained. There are no
flux variations over the two VPs in which the source has been detected (see Fig.1). The low upper limit $<4.2
\cdot 10^{-8} ~\cm2s$ derived in the VP=21.0 has $(TS)^{1/2} = 0$ and is not statistically significant.
 The EGRET source map is quite extended and round with a high value of $\theta_{95}= 1.13$ $\deg$.

 The cluster A388 falls within the $95 \%$ confidence level position error contour of the source (see Fig.4) and
 other Abell clusters are found in the field of this EGRET source. No other possible counterpart
 of the gamma-ray source is found in the field of this EGRET source.\\
 A388 has a redshift of $z =0.134$ and an estimated X-ray
luminosity of $L_{2-10 ~{\rm keV}} \simlt 3.47 \cdot 10^{44}$ erg s$^{-1}$, as reported in the HEAO-A1 all-sky
survey (Kowalski et al. 1984). No other relevant X-ray information is available for this cluster. There are two
NVSS radio sources correlated with this cluster with flux $S_{1.4} = 8.42 \pm 0.50$ and $4.03 \pm 0.48$ mJy,
respectively.\\
 Due to these evidence, this could be considered as a probable - but not yet definite - association
between galaxy clusters and EGRET gamma-ray source.

\subsection{3EG J0439+1105}

This EGRET source has been detected with a $(TS)^{1/2} = 4.2$ in the VP=P1234. It has a flux $F_{P1234}(>100 ~{\rm
MeV})=(9.4 \pm 2.4) \cdot 10^{-8} ~\cm2s$ with a power-law spectral index $\gamma = 2.44 \pm 0.29$. There are no
statistically significative flux variations over the different VPs in which the source has been detected (see
Fig.1) and the statistically significant upper limits are consistent with this conclusion.
 The EGRET source is quite extended with a high value of $\theta_{95}= 0.92$ $\deg$.

The cluster A497 falls within the $95 \%$ confidence level position error contour of the source. No other
gamma-ray source counterpart is found in the field of this EGRET source (see Fig.5).\\
 A497 has an
estimated redshift of $z \sim 0.14$ and an estimated X-ray luminosity of $L_{2-10 ~{\rm keV}} \sim 3.89 \cdot
10^{44}$ erg s$^{-1}$ (Ulmer et al. 1980). There is a NVSS radio source correlated with the cluster with a flux
$S_{1.4} = 3.27 \pm 0.47$ mJy.\\
 Due to these evidence, this could be considered as a probable - but not yet definite - association between
galaxy clusters and EGRET gamma-ray source.

\subsection{3EG J0215+1123}

This EGRET source has been detected with a $(TS)^{1/2} = 4.4$ in the VP=21.0. It has a flux $F_{21.0}(>100 ~{\rm
MeV})=(18.0 \pm 5.0) \cdot 10^{-8} ~\cm2s$ with a quite low power-law spectral index $\gamma = 2.03 \pm 0.62$.
There are no other definite detection of this EGRET source over other viewing periods (see Fig.1).
 The upper limit of  $6.0 \cdot 10^{-8} ~\cm2s$ obtained in the VP=317.0 is not statistically significant since it has $(TS)^{1/2} = 0$.
 The EGRET source is quite extended with a high value of $\theta_{95}= 1.06$ $\deg$ and is elongated in the
south-north direction.

The cluster A331 falls near the center of the EGRET source (see Fig. 5) and there is no other gamma-ray source
counterpart in the field of 3EG J0215+1123. There is another Abell cluster (A330) which is found at $\sim 1.1$
$\deg$ south of A331. The cluster A331 has a redshift of $z = 0.186$ and an X-ray luminosity of $L_{2-10 ~{\rm
keV}} \simlt 5.01 \cdot 10^{44}$ erg s$^{-1}$ as estimated in the HEAO-A1 all-sky survey (Kowalski et al. 1984).
There are four NVSS radio sources which are correlated with this cluster and their flux is $S_{1.4}= 3.68 \pm
0.48; 3.38 \pm 0.48; 4.73 \pm 0.49; 133.93 \pm 4.42$ mJy, respectively.\\
 Due to these evidence, this could be considered as a
probable - but still questionable - association between galaxy clusters and EGRET gamma-ray source.

\subsection{3EG J2248+1745}

This EGRET source has been detected with a $(TS)^{1/2} = 4.1$ in the VP=P1234. It has a flux $F_{P1234}(>100 ~{\rm
MeV}) =(12.9 \pm 3.5) \cdot 10^{-8} ~\cm2s$ with a power-law spectral index $\gamma = 2.11 \pm 0.39$. The flux
variations over the different VPs are at less than the $2 \sigma$ level, but the behaviour of this source is quite
different from the stationary ones in Fig.1.
 The EGRET source is quite extended with a value $\theta_{95}= 0.94$ $\deg$ and is elongated in the south-north direction.

The cluster A2486 falls within the $95 \%$ confidence level position error contour of the source. No other
gamma-ray source counterpart is found in the field of this EGRET source.\\
 A2486 has a redshift of $z = 0.143$ and an  X-ray luminosity of
$L_{2-10 ~{\rm keV}} \simlt 1.48 \cdot 10^{44}$ erg s$^{-1}$ as estimated in the HEAO-A1 all-sky survey (Kowalski
et al. 1984). There are two NVSS radio sources which are correlated with this cluster and their flux is $S_{1.4} =
4.60 \pm 0.51$ and $4.84 \pm 0.49$ mJy, respectively.\\
 Due to these evidence, and in particular to the flux variations
over the various VPs, this case should not  be considered as a possible association between galaxy clusters and an
EGRET gamma-ray source.

\subsection{3EG J1212+2304}

This EGRET source has been detected with a $(TS)^{1/2} = 3.3$ in the VP=Virgo2. It has a flux $F_{Virgo2}(>100
~{\rm MeV}) =(19.7 \pm 7.7) \cdot 10^{-8} ~\cm2s$ with a power-law spectral index $\gamma = 2.76 \pm 0.60$. Even
though the flux of this source may be affected by confusion, there are quite strong  flux variations over the
three different VPs in which this source has been detected (see Fig.1). The EGRET source is quite extended with a
value $\theta_{95}= 0.88$ $\deg$ and is elongated in the south-north direction.

The cluster A1494 falls within the $95 \%$ confidence level position error contour of the source. No other
gamma-ray source counterpart is found in the field of this EGRET source.\\
  A1494 has a redshift of $z=0.159$ and an X-ray luminosity of $L_{2-10 ~{\rm keV}} \simlt 1.95 \cdot 10^{45}$ erg s$^{-1}$ as estimated
  in the HEAO-A1 all-sky survey (Kowalski et al. 1984). There are five NVSS radio sources which are correlated with this cluster and their
flux is $S_{1.4} = 4.91 \pm 0.48; 8.05 \pm 0.49; 23.68 \pm 1.74; 17.53 \pm 0.66; 5.15 \pm 0.47 $ mJy,
respectively.\\
 Due to these evidence, and in particular to the flux variations over the various VPs, this case
should not  be considered as a possible association between galaxy clusters and EGRET gamma-ray source.

\subsection{3EG J1347+2932}

This EGRET source has been detected with a $(TS)^{1/2} = 4.0$ in the VP=P1234. It has a flux $F_{P1234}(>100 ~{\rm
MeV})=(9.6 \pm 2.9) \cdot 10^{-8} ~\cm2s $ with a power-law spectral index $\gamma = 2.51 \pm 0.61$.  There are no
strong  flux variations over the different VPs in which this source has been detected (see Fig.1), even though the
flux of this source may be affected by confusion. However, we noticed that there is only one independent detection
of this source in the VP=4.0 which does not allow to draw any definite conclusion on its possible variability.
 The EGRET source is quite extended with a value of $\theta_{95}= 0.95$ $\deg$ and is irregular.

The cluster A1781 falls within the $95 \%$ confidence level position error contour of the source. No other
gamma-ray source counterpart is found in the field of this EGRET source.\\
  A1781 has a
redshift of $z = 0.0618$ and an X-ray luminosity of $L_{2-10 ~{\rm keV}} \sim 1.15 \cdot 10^{44}$ erg s$^{-1}$ as
estimated in the HEAO-A1 all-sky survey (Kowalski et al. 1984). There is one radio galaxy ({\it FIRST
J134159.7+294653}) at a redshift of $z = 0.0457$ which is associated with the cluster. There is no evidence for
NVSS radio sources correlated with A1781.\\
 Due to these evidence, and in particular to the uncertainties in the flux variations over the various VPs,
this case should not  be considered as a possible association between galaxy clusters and EGRET gamma-ray source.

\subsection{3EG J1424+3734}

This EGRET source has been detected with a $(TS)^{1/2} = 4.4$ in the VP=P1. It has a flux $F_{P1}(>100 ~{\rm
MeV})=(16.3 \pm 4.9)\cdot 10^{-8} ~\cm2s $ with a power-law spectral index $\gamma = 3.25 \pm 0.46$. There are no
strong flux variations over the different VPs in which this source has been detected (see Fig.1) and the lowest
upper limit of $< 16.1\cdot 10^{-8} ~\cm2s $ found in the VP=201.+ is not statistically significant because it has
$(TS)^{1/2}=0$. The EGRET source is quite regular with a value $\theta_{95}= 0.88$ $\deg$ and with an emission
tail in the north-west side of the field.

Two rich clusters (A1914 and A1902) fall within the $95 \%$ confidence level position error contour of the source
(see Fig.4) . No other gamma-ray source counterpart is found in the field of this EGRET source.\\
 A1902 has is a redshift of $z = 0.16$ and  is associated with the X-ray source RXJ1421.6+3717
(Boehringer et al. 2000) with an X-ray  flux $F_{0.1-2.4 ~{\rm keV}} = (4.3 \pm 10.8)\cdot 10^{-12} ~{\rm
erg}~{\rm cm}^{-2}~{\rm s}^{-1}$ and a luminosity $L_{0.1-2.4 ~{\rm keV}} \sim 6\cdot 10^{44} ~{\rm erg}~{\rm
s}^{-1}$. Its X-ray luminosity has been also estimated to be $L_{2-10 ~{\rm keV}} \sim 1.26 \cdot 10^{44}$ erg
s$^{-1}$ in the HEAO-A1 all-sky survey (Kowalski et al. 1984). There is one radio source ({\it FIRST
J142140.4+371731}) associated to the cluster galaxy {\it MAPS-NGP $0-272-0323568$} found at a redshift of $z =
0.16$. There is also a NVSS radio source with flux $S_{1.4} = 3.51 \pm 0.46$ mJy which is associated with the
cluster.\\
 A1914 has a redshift of $z = 0.1712$ and
is associated to the X-ray source RXJ1426.0+3749 with an X-ray flux $F_{0.1-2.4 ~{\rm keV}} = (12.90 \pm 5.2)\cdot
10^{-12} ~{\rm erg} ~{\rm cm}^{-2} ~{\rm s}^{-1}$. Its X-ray luminosity has been estimated to be $L_{0.1-2.4 ~{\rm
keV}} \sim 15.91 \cdot 10^{44}$ erg s$^{-1}$ (Boehringer et al. 2000). Three NVSS radio sources are correlated
with A1914 and they have flux $S_{1.4} = 9.91 \pm 1.50, 30.86 \pm 1.77, 20.75 \pm 4.08 $ mJy, respectively. The
cluster A1914 also hosts the Very Steep Spectrum radio galaxy {\it 1474+380 (4C 38.39)} (Komissarov \& Gubanov
1994). This cluster has also a bright radio halo (see Fig.7) detected in the VLA with a flux of $S_{1.4} = 50$ mJy
and a power $P_{1.4} = 6.31\cdot 10^{24}$ W Hz$^{-1}$ (Giovannini \& Feretti 2000).\\
 Due to the previous evidence we consider this case as a probable
candidate for the correlation of galaxy clusters and EGRET unidentified gamma-ray sources.
\begin{figure}[tbh]
\begin{center}
  \parbox{15.cm}{\fbox{a1914fig7.JPG}}
  \vskip 15.truecm
\end{center}
\caption{\footnotesize {One of the most probable associations between galaxy clusters and EGRET unidentified
gamma-ray sources: A1914. This cluster has a radio halo/relic inhabiting the cluster and a number of identified
radio galaxies in the ICM. Shown are the EGRET map image with the cluster X-ray brightness contours (right), the
ROSAT-HRI X-ray image of the cluster (center) with the NVSS radio sources in the field (red circles) and the radio
halo/relic image observed with the VLA at $1.4$ GHz (left). }} \label{figure: 7}
\end{figure}

\subsection{3EG J1337+5029}

This EGRET source has been detected with a $(TS)^{1/2} = 4.4$ in the VP=P1234. It has a flux $F_{P1234}(>100 ~{\rm
MeV})=(9.2 \pm 2.6) \cdot 10^{-8} ~\cm2s$ with a power-law spectral index $\gamma = 1.83 \pm 0.29$. There are no
strong flux variations over the different VPs in which this source has been detected (see Fig.1). The EGRET source
is quite regular with a value $\theta_{95}= 0.72$ $\deg$.

The rich cluster A1758  falls within the $95 \%$ confidence level position error contour of the source, very close
to the center of the EGRET source map (see Fig.5). No other gamma-ray source counterpart is found in the field of
this EGRET source.\\
 A1758 is the most distant cluster listed in Table 1. It has a redshift of $z =
0.279$ and is associated with the X-ray source RXJ1332.7+5032 with an X-ray flux $F_{0.1-2.4 ~{\rm keV}} = (5.6
\pm 9.5) \cdot 10^{-12} ~{\rm erg} ~{\rm cm}^{-2} ~{\rm s}^{-1}$ and a luminosity $L_{0.1-2.4 ~{\rm keV}} \sim 1.8
\cdot 10^{45} ~{\rm erg} ~{\rm s}^{-1}$ (Boehringer et al. 2000). The ROSAT-PSPC observation yielded a temperature
$kT \approx 4.1$ keV which is found to be much lower than the ASCA (SIS+GIS) temperature of $T \approx 9.33$ keV
(Rizza et al. 1995). An X-ray luminosity of $L_{2-10 ~{\rm keV}} \approx (1.43 \pm 0.06) \cdot 10^{45}$ erg
s$^{-1}$ has been estimated independently by Wu et al. (1999).\\
 There are four NVSS radio sources correlated with this cluster with a flux $S_{1.4} = 109.83 \pm 3.86, 8.69 \pm
1.42, 14.83 \pm 1.51, 5.18 \pm 0.43 $ mJy, respectively. A1758 also hosts the narrow tailed radio galaxy {\it
87GB~133050.3+504752} (Feretti et al. 1992).
 This cluster also shows a diffuse radio emission (see Fig.8) which could be possibly identified with an extended radio halo
 (Giovannini \& Feretti 2000).\\
 Due to the previous evidence we consider this case as a probable candidate for the correlation of galaxy clusters
and EGRET unidentified gamma-ray sources.
\begin{figure}[tbh]
\begin{center}
  \parbox{15.cm}{\fbox{a1758fig8.JPG}}
  \vskip 15.truecm
\end{center}
\caption{\footnotesize {One of the most probable associations between galaxy clusters and EGRET unidentified
gamma-ray sources: A1758. This cluster has a radio halo/relic inhabiting the cluster and a number of identified
radio galaxies in the ICM. Shown are the EGRET image with the cluster X-ray brightness contours superposed
(right), the ROSAT-HRI X-ray image of the cluster (center) with the NVSS radio sources in the field (red circles)
and the radio halo/relic image observed with the VLA at $1.4$ GHz (left). }} \label{figure: 8}
\end{figure}

\subsection{3EG J1447-3936}

This EGRET source has been detected with a $(TS)^{1/2} = 4.5$ in the VP=P1234. It has a flux $F_{P1234}(>100 ~{\rm
MeV}) =(11.0 \pm 2.7) \cdot 10^{-8} ~\cm2s$ with a power-law spectral index $\gamma = 2.45 \pm 0.34$. There are no
strong flux variations over the different VPs in which this source has been detected (see Fig.1). However, the
only two independent detections of this source do not allow to draw any definite conclusion on its variability.
 The EGRET source is quite regular with a value $\theta_{95}= 0.87$ $\deg$.\\
 The cluster A774  falls within the $95 \%$ confidence level position error contour of the source. No other
gamma-ray source is found in the field of this EGRET source.\\
 A774 is a poor cluster with very limited information at other wavelengths.
There is no radio source correlated with this cluster.\\ We thus believe that this case of spatial correlation is
likely due to projection effects.

\section{The gamma-ray -- radio  correlation}

Many of the clusters listed in Table 1 (namely, 17 out of 24 clusters) have also bright NVSS radio sources within
their Abell radius ($ \approx 3 h_{50}^{-1}$ Mpc, of order of the virial radius), six clusters have identified
bright radio galaxies in their environment and three clusters (A1758, A1914 and A85) have also a radio halo or
radio relic inhabiting their ICM (see Figs. 6-8). Since the EGRET sources have been selected to be of high
galactic latitude, the NVSS radio sources are very likely non-identified (active) radio galaxies, as also
indicated by the NVSS-NRAO images of many of the radio sources found in our analysis of the specific EGRET sources
counterparts as discussed in Sect.3 above.

The 9 EGRET sources in Table 1 which are marked with an asterisk are those more likely associated to galaxy
clusters according to our analysis of the specific sources presented in Sect.3 above. These galaxy clusters are
quite peculiar since all of them have bright NVSS radio sources in their environment and six of them have also
bright radio galaxies living in their environment. Three of the clusters which are more probably associated with
these EGRET sources show also the presence of extended radio halos or relics. Hence, such galaxy clusters which
have strong radio emission (either diffuse or associated with member galaxies) show the direct presence of a
population of relativistic electrons which are injected in their ICM.\\
 Radio galaxies, which are mainly found in the central regions of the clusters, may inject into the cluster ICM
 large quantities of energy transported by their relativistic jets.
 This energy is probably changed from an electromagnetic form to a pair plasma, to an ion plasma and (at least
 partially along the way) into energetic photons (see, e.g., Blandford 2001). Such high-energy particles and
 photons may produce $\sim$ GeV gamma-ray emission which can be observed by EGRET.
 The active galaxy radio power correlates with the gamma-ray power (Padovani et al. 1993) indicating that
 radio louder galaxies emit more gamma-ray power which, in turn, seems to be associated with relativistic beaming
 of the jets (see, e.g., Urry \& Padovani 1995).
 Also the particles injected into the ICM by the radio-galaxy jets may diffuse in the magnetized ICM (Colafrancesco
 \& Blasi 1998) and interact with the ICM particles (mainly electrons and protons)
to produce diffuse radio emission (Blasi \& Colafrancesco 1999), heating of the ICM itself (Yamada \& Fujita 2001,
Kaiser \& Alexander 1999, Inoue \& Sasaki 2001, Nath \& Roychowdhury 2002) and secondarily produced gamma-ray
emission (Colafrancesco \& Blasi 1998, Blasi 2000).\\
 In addition, particles in the ICM could be efficiently accelerated at the accretion shocks located at the cluster
 periphery as well as at the ICM shocks produced by subcluster mergings (see, e.g.,
 Miniati et al. 2000) and/or by fast galaxy motions.
 The subsequent interaction of the accelerated particles with the surrounding hot, magnetized
  ICM can again produce diffuse radio halo/relic emission (see, e.g., Blasi \& Colafrancesco 1999, Sarazin 2001) and diffuse
  gamma-ray emission at $E > 100 ~{\rm MeV}$ (Colafrancesco \& Blasi 1998).
  On top of these acceleration mechanisms, it has been recently shown that dark matter particle (neutralinos)
annihilation -- a mechanism which is especially efficient  in the central regions of the clusters -- may produce
both diffuse radio halo emission and diffuse gamma-ray emission visible at $E > 100 ~{\rm MeV}$ (Colafrancesco \&
Mele 2001).

 The presence
of such relativistic particles into the ICM strongly suggests, in conclusion, that themselves and/or their parent
population (e.g., relativistic protons, dark matter particles) can be responsible for a substantial gamma-ray flux
at the EGRET energies ($E > 100$ {\rm MeV}) as well as non-thermal radio emission through different mechanisms.
 Thus, we propose that there should be a close connection between radio emission (either diffuse or associated
 with individual active galaxies) and gamma-ray emission in galaxy clusters.
\begin{figure}[tbh]
\begin{center}
\epsfig{figure=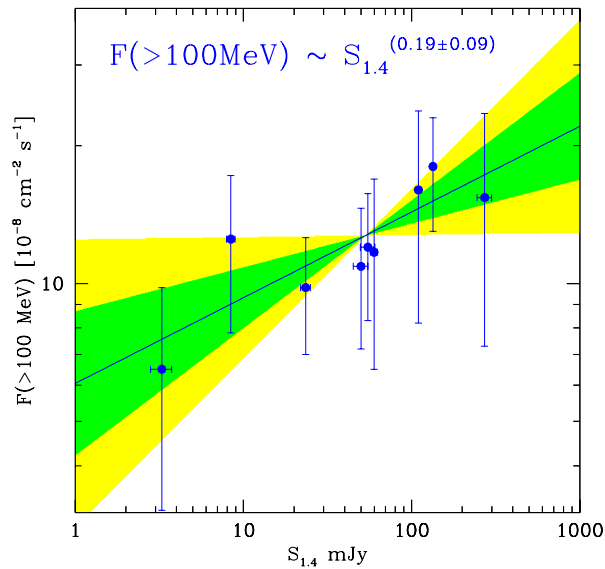,width=8.cm,angle=0.}
\end{center}
\caption{\footnotesize { The correlation between the gamma-ray flux, $F(>100 ~{\rm MeV})$, and the radio flux at
$1.4$ GHx, $S_{1.4}$, shown by the EGRET source -- cluster associations listed with an asterisk in Table 1.
 The best fit curve $F(>100 ~{\rm MeV}) \sim S^{0.19}_{1.4}$ is shown (solid) together with the $1 \sigma$
(green/dark gray) and $2 \sigma$ (yellow/pale gray) confidence level regions of the fit. }}
 \label{figure: 9}
\end{figure}

Based on the previous arguments, we should expect a positive correlation between the flux of the radio sources
associated with the galaxy cluster and the gamma-ray flux of the relative EGRET source. In fact, we found a
positive correlation between the radio flux at $1.4$ GHz, $S_{1.4}$, of the brightest radio source in the cluster
and the EGRET source flux, $F(>100 ~{\rm MeV})$, which is reported in the Third EGRET Catalog (Hartman et al.
1999). Specifically, we find a correlation $F(>100 ~{\rm MeV})= A S_{1.4}^B$ with $A= 6.053^{+2.637}_{-1.836}$ and
$B = 0.187 \pm 0.091$ ($1 \sigma$ errors) using gamma-ray fluxes selected in the different observing periods of
the EGRET source (see Fig.9). The best fit has a $\chi^2=2.38$ which gives a probability $P=0.064$ for the null
hypothesis of a random distribution for the radio and gamma-ray flux of the nine sources in our analysis. This
gives a statistical confidence level of $\approx 2.05 \sigma$. We show in Fig.9 the best fit curve and the $1
\sigma$ and $2 \sigma$ confidence level regions for the fit.
 A similar result obtains considering the
total radio flux from the clusters (most of the clusters here considered have more than one radio source in their
environment) associated with the previous EGRET sources. The reason for such similar result is that in many cases
the cluster radio flux at $1.4$ GHz is dominated by the brightest radio source in the cluster.

Even though the large uncertainties in the EGRET source fluxes do not allow to draw any strong conclusion for the
universality of such correlation, the present results indicate that there is a connection between the activity of
the cluster ICM, and of its active galaxy content, and the overall gamma-ray behaviour of these large scale
structures, an indication that can be definitely confirmed by the next generation gamma-ray telescopes. The
detailed follow-up of the galaxy populations of the clusters most probably associated to the 9 EGRET sources here
selected  is an important aspect in this research field but it is far beyond the aims of the present work and will
be tackled in a forthcoming paper.

\section{The gamma-ray -- X-ray luminosity correlation}

To further strengthen our argument presented in the previous sections,  we looked for other intrinsic correlations
among galaxy clusters and unidentified EGRET sources. Specifically, we correlated the cluster X-ray luminosity,
$L_{\rm X}$, with the luminosity of the associated EGRET source under the assumption that it is physically
associated to the cluster and, hence, has the same redshift. We derived the gamma-ray luminosity, $L_{\gamma}$, of
each EGRET source from the gamma-ray fluxes at $E > 100$ {\rm MeV} given in the Third EGRET catalog (Hartman et
al. 1999) using the cluster optical redshift given in Table 1.
We consider the same EGRET-cluster associations (with the exception of the source 3EGJ0159-3603 associated with
the clusters A219 and A2963 because no reliable redshift is available for these clusters) which show the $F(>100
~{\rm MeV}) - S_{1.4}$ correlation analyzed in the previous Sect.4.
The $L_{\gamma} - L_{\rm X}$ correlation shown by the data (see Fig.10) is fitted by $L_{\gamma} = C L_{\rm X}^D$
with best fit values $C= 0.06 \pm 0.01$ and $D= 0.593 \pm 0.122$ ($1 \sigma$ errors). A similar result, however,
is also found for viewing periods P1234 which do not show, in general, the best S/N ratios for the detected EGRET
sources: in this last case we found $C= 0.08 \pm 0.02$ and $b= 0.328 \pm 0.120$ ($1 \sigma$ errors). In any case,
the $L_{\gamma} - L_{\rm X}$ relation shown by the data is significant at more than $2.7 \sigma$ confidence level.

Such a $L_{\gamma}-L_{\rm X}$ correlation indicates a connection between the physical status of the cluster ICM,
and of its galaxy content, and the overall gamma-ray emissivity of the cluster: such a connection is indeed
expected in the viable model for the gamma-ray emission of galaxy clusters. In fact, both the diffuse emission
arising from the interaction of relativistic particles with the cluster ICM and the one arising from a
superposition of the gamma-ray emission associated with individual galaxies within the cluster predict a relation
$L_{\gamma} \sim L_{\rm X}^{a}$ with $a \approx 0.45 - 0.85$. Specifically, the cluster gamma-ray luminosity
produced by non-thermal electron bremsstrahlung (see, e.g., Longair 1993),
 \be
L_{\gamma} \propto n_{{\rm e, rel.}} n R^3 ~,
 \ee
 that produced by $\pi^0 \to \gamma + \gamma$ in $pp$ collisions
(Colafrancesco \& Blasi 1998),
 \be
  L_{\gamma} \propto n_{{\rm p, rel.}} n R^2 ~,
  \ee
 and the one produced  by $\pi^0 \to \gamma + \gamma$ in dark matter annihilation (Colafrancesco \& Mele 2001),
 \be
 L_{\gamma} \propto n^2 R^3 ~,
 \ee
 naturally
correlate with the cluster X-ray luminosity mainly given by thermal bremsstrahlung,
\be
L_{\rm X} \propto n^2 T^{1/2} R^3 ~,
 \ee
 through their dependence from the ICM particle density, $n$. Here, the densities
of relativistic electrons, $n_{\rm e, rel.}$, and relativistic protons, $n_{\rm p, rel.}$, are decoupled from the
ICM density while the dark matter density is proportional to the ICM density. Note that  a scaling similar to that
in Eq.(3) applies also to the gamma-ray emission arising from the superposition of the cluster galaxies. Using the
previous scalings and the observed X-ray luminosity -- temperature relation, $L_{\rm X} \sim T^b$ with $b \approx
3$ (see, e.g., Arnaud \& Evrard 1999, Wu et al. 1999),
 a correlation
$L_{\gamma} \sim L_{\rm X}^{a}$, with $a \approx 0.45 - 0.85$  is predicted by the previous models, in agreement
with our results shown in Fig.10.
\begin{figure}[tbh]
\begin{center}
\epsfig{figure=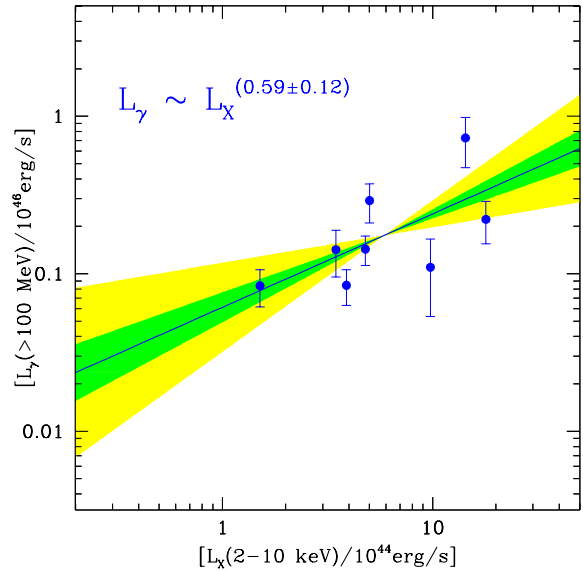,width=8.cm,angle=0.}
\end{center}
\caption{\footnotesize { The $L_{\gamma} - L_{\rm X}$ correlation shown by the clusters listed with an asterisk in
Table 1. The best fit curve (solid line) is shown together with the $1 \sigma$ (green/dark-gray area) and $3
\sigma$ (yellow/pale-gray area) confidence level region for the fitting parameters. The X-ray luminosity, in units
of $10^{44}$ erg s$^{-1}$,  is given in the $2-10$ keV energy range and the gamma-ray luminosity, at $E> 100$ {\rm
MeV}, of the associated EGRET source is given in units of $10^{46}$ erg s$^{-1}$. }} \label{figure: 5}
\end{figure}

\section{Present conclusions and future perspectives}

In this paper we reported the preliminary evidence for an association of galaxy clusters with unidentified, high
galactic latitude ($|b|>20$ $\deg$) gamma-ray sources in the Third EGRET catalog. Our selection criteria
eventually allowed us to identify 9 EGRET sources most probably associated to 12 galaxy clusters (see the sources
marked with an asterisk in Table 1) which have the following characteristics:
 {\it i)} the clusters are found within the $95 \%$ confidence level position error contours of the relative EGRET source
 map for which there is no other known counterpart;
 {\it ii)} the selected EGRET sources have flux  $F(>100 ~{\rm MeV}) \simlt 2 \cdot 10^{-7} \cm2s$ and flux
 variability $\simlt 20 \%$ over their viewing periods;
 {\it iii)} the gamma-ray spectral index of the EGRET source are found in the range $\approx 2 - 3$;
 {\it iv)} the 12 galaxy clusters which are most probably associated with the previous 9 unidentified
 EGRET sources have bright radio sources (identified radio galaxies,
radio halo/relic, NVSS bright radio source) in the clusters environment;
 {\it v)} the nine EGRET sources selected according to the previous criteria show a
 correlation $F(>100 ~{\rm MeV}) \sim S_{1.4}^{0.19 \pm 0.09}$ between their gamma-ray flux,
 $F(>100 ~{\rm MeV})$, and the radio flux at $1.4$ GHz, $S_{1.4}$, of the brightest radio source in the associated clusters;
 {\it vi)} the same EGRET sources and the same clusters show also a correlation,
 $L_{\gamma} \propto L_{\rm X}^{0.59 \pm 0.12}$ between the gamma-ray luminosity at $E> 100$ ~{\rm MeV}, $L_{\gamma}$, of the EGRET
 source and the X-ray luminosity, $L_{\rm X}$, of the associated clusters.

 From our analysis of the sample listed in Table 1, we expected {\it a priori} a spatial
correlation between unidentified EGRET sources and galaxy clusters at the $\sim 1.73 \sigma$ confidence level (see
Sect.2). We noticed, however, that this should be considered as a lower limit to the true statistical confidence
level of the correlation  since the effect of the non-uniform EGRET sky coverage has to be taken into account and
it would tend to increase the statistical significance level of the EGRET-cluster spatial correlation (see Sect.2
for a discussion). The detailed analysis (see Sect.3) of each specific EGRET source yielded a most probable
spatial association between 9 EGRET sources and 12 Abell clusters selected from the original list
 of 18 EGRET sources associated with 24 clusters given in Table 1: such a spatial
 correlation  is found at $\sim 3 \sigma$ confidence level and might decrease to
 $\sim 2.5 \sigma $ eliminating the still questionable case of the spatial association between A1688 and 3EGJ1310-0517
 (see Fig.4).
Note, again, that this is a lower limit to the true statistical confidence level of the correlation because of the
effect of the non-uniform EGRET sky coverage.\\
 The gamma-ray--radio correlation found for the nine most probable EGRET-cluster associations,
 \be
F(>100 ~{\rm MeV}) \sim S_{1.4}^{0.19 \pm 0.09}~,
 \ee
 is at $\approx 2.05 \sigma$ confidence level (we
considered here only the statistical uncertainties).
 The gamma-ray -- X-ray correlation shown by the same EGRET-cluster associations,
 \be
 L_{\gamma} \propto L_{\rm X}^{0.59 \pm 0.12}~,
 \ee
 is at $\approx 4.9 \sigma$ confidence level (again, we
considered only the statistical uncertainties).

We estimated the diffuse gamma-ray fluxes predicted in the available models under reasonable assumption for the
energy density of relativistic particles in the ICM (see Sects. 4 and 5 above) for the galaxy clusters listed in
Table 1 and we found that the total diffuse fluxes are usually a factor 2-4 below the fluxes actually detected for
the associated EGRET sources. So, to recover the gamma-ray flux of the EGRET sources we have to consider that, at
least, a comparable fraction of the cluster gamma-ray flux is contributed also by the (active) radio galaxies
living within the cluster. We found, consistently with such a picture, that all of the clusters which are probable
counterparts of the unidentified EGRET sources host several bright radio galaxies in their environment. Such radio
galaxies can be, or have recently passed through a phase of substantial gamma-ray emission, according to the
leading unified scheme scenarios for radio galaxy evolution (see, e.g., Urry \& Padovani 1995).
 Thus, the EGRET data require that the gamma-ray flux associated to the
relative galaxy clusters are likely due to a superposition of diffuse and concentrated gamma-ray emission.

The low flux variability of the associated EGRET sources does not indicate a strong contamination from very bright
[$F(>100~ {\rm MeV}) > 5 \cdot 10^{-7} ~\cm2s$] AGN-like gamma-ray sources with strong flux variability. This is
clearly shown by the comparison of the flux changes for the EGRET sources more probably associated with clusters
(see Fig.1) with the flux changes of the EGRET sources spatially correlated with clusters and whose gamma-ray
emission is dominated by bright AGNs (see Fig.2).
\begin{figure}[tbh]
\begin{center}
\epsfig{figure=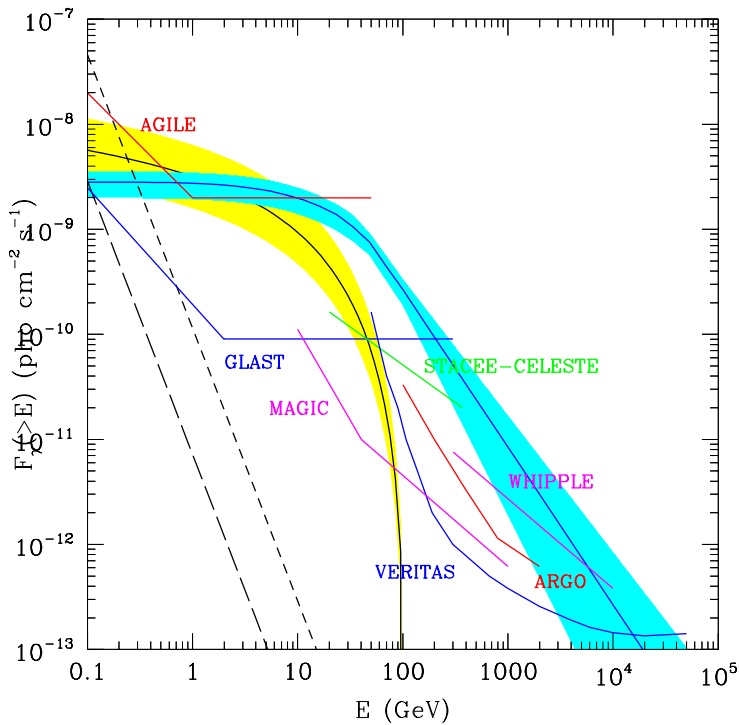,height=10.cm,angle=0.}
\end{center} \caption{\footnotesize {Theoretical  predictions for the gamma-ray flux $F_{\gamma}(>100 ~{\rm MeV})$
expected for a Coma-like cluster are shown as a function of the gamma-ray energy and are compared with the
sensitivity of the next generation space-borne  and ground-based gamma-ray experiments: non-thermal electron
bremsstrahlung (Sreekumar et al. 1996, Colafrancesco 2001b) for two choices of the intracluster magnetic field ($B
= 0.3 \mu$G: short-dashed curve and $B=1 \mu$G: long-dashed curve); decay of neutral pions produced in $pp$
collisions (Colafrancesco \& Blasi 1998) (blue curve and the associated theoretical uncertainties given in the
cyan region); decay of neutral pions produced in the annihilation of dark matter neutralinos (Colafrancesco \&
Mele 2001) (black solid curve and the associated theoretical uncertainties given in the yellow region). Due to the
very different spatial resolution of the various experiments reported, we show here the case of their sensitivity
for point-like sources.}}
 \label{figure: 11}
\end{figure}

The spectral indices of the most probable EGRET-cluster associations are found to be in the range $\approx 2 -
3.5$, values which are consistent with the expectations from model of the diffusion of relativistic particles in
the ICM, and seem to be quite larger than the very flat spectral indices ($\gamma \simlt 2$) shown by the EGRET
sources associated with pulsars. Theoretical models for cluster gamma-ray emission predict in fact slopes in the
range $\gamma \sim 1.8 - 3.2$ (see, e.g., Fig.11, see also Blasi 2000), going from annihilation of dark matter
neutralinos to non-thermal electron bremsstrahlung. Active galaxies with a substantial gamma-ray emission at the
flux level shown by the EGRET-cluster associations also have spectral slopes $\gamma \sim 2 - 2.8$, as shown in
Fig.3 (see also Hartman et al. 1999). Thus, the superposition of gamma-ray emission of both diffuse origin and
coming from the active galaxies shall certainly show overall spectral indices which are consistent with those of
the nine EGRET sources selected in our analysis.

In conclusion, we found that there are several converging evidence (even though still preliminary) of an
association between unidentified EGRET sources at high galactic latitude ($|b|>20$ $\deg$) and galaxy clusters
which show an enhanced radio activity in their ICM as triggered by radio (or active) galaxies or by non-thermal
phenomena giving rise also to radio halos and relics (see, e.g., Colafrancesco 2001a,b). These evidence are found
at several levels, from the geometrical spatial association with a minimal statistical confidence level of $\sim
2.5 \sigma$ (see Sect.2), to the gamma-ray flux and luminosity correlations with the radio and X-ray data of the
associated clusters with a statistical confidence level of $\sim 2.1\sigma$ and $\sim 4.9 \sigma$, respectively
(see Sect.4 and 5).

Even though the cluster sample we derived here is far from being an {\it a priori} flux limited sample, the
correlation we found with unidentified EGRET gamma-ray sources can be considered as the first evidence of the
expected distribution of the gamma-ray luminosity of ``active'' galaxy clusters.

There have been recently other attempts to investigate the possible association of galaxy clusters with EGRET
gamma-ray sources. In fact, Totani \& Kitayama (2000, hereafter TK) proposed that only galaxy clusters which are
just dynamically forming might be bright sources of gamma-rays due to Inverse Compton Scattering (ICS) of CMB
photons by high-energy electrons accelerated at the shock waves induced by gravity during the early formation of
large scale structures. Their model would predict, for instance, a gamma-ray flux of $F(>100 ~{\rm MeV}) \sim 6.5
\cdot 10^{-7} ~\cm2s$ for a Coma-like cluster undergoing a merger event [roughly a factor 16 higher than the
actual upper limit, $F(>100 ~{\rm MeV}) \sim 4 \cdot 10^{-8} ~\cm2s$ found for Coma in the EGRET database (see,
e.g., Sreekumar et al 1996)]. As a consequence, TK predicted that a few tens of clusters ($\sim$ 20 to 50 with a
limiting flux $F(>100 ~{\rm MeV}) \sim 10^{-7} ~\cm2s$) should have already been detected by EGRET. The absence of
any correlation between the ROSAT Bright Cluster Sample (Ebeling et al. 1998) or the ACO (Abell et al. 1989)
cluster catalog and the EGRET source catalog should depend, according to TK, on the large extension of these
``just forming clusters'' which would cause a huge dimming of their X-ray surface brightness as well as of their
surface number density of galaxies in the optical with respect to the population of virialized, relaxed
clusters.\\
 However, more recently and after the submission of our paper, the same authors (Kawasaki \& Totani 2001, hereafter KT) found instead a
 strong correlation between merging clusters and steady unidentified EGRET sources at high galactic
 latitude ($|b| > 45$ $\deg$). They used the same data sets (Third EGRET catalog and ACO cluster catalog) and found that 9 close
 pair/groups (CPG) of Abell clusters have a significative statistical level of spatial association with 7 steady
 unidentified EGRET sources.
Interestingly, 6 out of these 7 EGRET sources are coincident with the EGRET-cluster associations found in our
analysis (see Table 1) while the last case (the clusters A1564 and A1581 associated with 3EGJ1235+0233) is not
found in our analysis because these clusters are found outside the $95 \%$ confidence level position error
contours of the relative EGRET source (see Sect.2).
 These last authors, nonetheless, suggested that the gamma-ray emission comes only from just forming/merging clusters
 with large, violent shocks, but not from usual ones in dynamically quiet regime where the violent shock has subsided.
They further concluded that their finding ``implies that the bulk of the steady unidentified EGRET sources in the
high latitude originate from forming clusters'' and ``indirectly give support to the gamma-ray cluster
hypothesis'' delineated in TK.

Let us briefly comment on this point. We notice here that since the gamma-ray clusters considered in TK and KT are
physically the same ``forming/merging clusters'' (their gamma-ray fluxes are evaluated according to the same ICS
model) and since TK predicted that a few tens of these clusters should have already been  detected by EGRET, there
seem to be a missing gamma-ray cluster problem in their approach because KT do not find the remaining
($\sim$13-43) bright gamma-ray clusters, as predicted by TK. A possible solution to this problem could be that the
large majority of the forming clusters are not ``just forming'' as suggested by TK but have experienced a strong
merging event more than a few Gyrs ago, so that the gamma-ray emission from the once accelerated primary electrons
has faded away due to their rapid energy losses ($t_{\rm cool} \simgt  2 \cdot 10^6 $ yr; see, e.g., Blasi 2000
and TK). The only gamma-ray clusters still remaining should be those which experienced a strong merger event in
the last $\sim 10^8$ yr.

But there are also other concerns as regards the energetics of the just forming/merging galaxy clusters. The
gamma-ray luminosity of the EGRET sources selected by KT are found in the range $L_{\gamma} \sim 10^{45} - 10^{46}
{\rm erg} ~{\rm s}^{-1}$ (see Fig. 10: note that most of the EGRET sources selected by KT are the same we select
in this paper) and should be emitted from primary electrons on a time scale $t_{\rm cool} \sim 2 \cdot 10^6$ yr.
This gamma-ray power should be compared with the total power, $L_{\rm merg} \sim E_{\rm merg}/ t_{\rm merg}$,
provided by the merging between two sub-cluster units with masses $M_1$ and $M_2$, respectively. Here the total
energy of the merger is $E_{\rm merg} \approx G M_1 M_2/d$ where $d$ is the typical sub-cluster separation at
which most of the energy is released on the time scale for the merging, $t_{\rm merg}\approx 10^9$ yr, of the
order of the crossing time for the considered cluster. Simulations show that equal-mass mergers are more effective
in releasing energy from its gravitational form to heating of the ICM and to particle acceleration at the ICM
shocks. Thus, the total power provided by the merger can be written as
 \be
L_{\rm merg} \approx 1.6 \cdot 10^{45} ~{\rm erg} ~{\rm s}^{-1} \bigg({M \over 10^{14} M_{\odot}} \bigg)^2
\bigg({d \over 1.5 Mpc}\bigg)^{-1}~, \label{eq_lmerg}
 \ee
(see also Blasi 2000). It is reasonable to consider that only a fraction $\varepsilon \sim 10^{-2}$ of the total
$E_{\rm merg}$ is transformed in particles which are shock-accelerated up to energies $E\simgt$ GeV while the bulk
of the total merging energy goes mainly into heating of the cluster ICM (see, e.g., Blasi 2000). Thus, the
gamma-ray luminosity emitted by primary electrons accelerated at the merging shocks can be written, in general, as
$L_{\gamma} \approx \varepsilon L_{\rm merg}$. The values of $L_{\gamma}$ of the  EGRET sources selected by KT
require, on average, an efficiency $\varepsilon \sim 1 - 10$ in the CGP clusters. This result seems to be strongly
in contrast with the available models for gamma-ray emission from a population of primary electrons for which
$\varepsilon \sim 10^{-2} - 10^{-1}$ is expected (Blasi 2000, Colafrancesco \& Blasi 1998, Sarazin 2001).\\
 Such a problem could be partially reduced if a substantial fraction of the gamma-ray emission from clusters is
 provided by $\pi^0 \to \gamma + \gamma$ decay  produced by $pp$
 interactions in the cluster ICM, where high-energy protons are accelerated at the same merging shocks but do not
appreciably loose their energy over an age comparable with $H_0^{-1}$ (see, e.g., Colafrancesco \& Blasi 1998).
 If a ratio $p/e^- \sim 10 - 100$ is assumed, then a large part of the cluster gamma-ray emission
could be dominated by $\pi^0$ decay and secondary electron emission by bremsstrahlung and ICS. This fact would
weaken the constraint $\varepsilon \sim 1-10$ for the values $L_{\gamma}$ of the EGRET sources produced by the
primary electrons in the approach of KT but would also provide a gamma-ray emission which is stationary with time
along the cluster lifetime (Blasi 2000). As a consequence, a large fraction of the $\sim$ 20-50 gamma-ray, merging
clusters predicted by TK should have been already detected by EGRET, which does not seem to be the case.\\
 On another side, if the gamma-ray luminosity of the EGRET source, $L_{\gamma}\sim \varepsilon
 L_{\rm merg} \approx 10^{45} ~{\rm erg} ~{\rm s}^{-1}$
 is provided by primary electrons accelerated at the merging shock, then a much larger energy amount, $E_{\rm merg}
 \sim E_{\gamma}/\varepsilon$, where $E_{\gamma} \approx L_{\gamma} \cdot t_{\rm cool} \sim 6 \cdot 10^{60-61}$ erg
 (we assume here $t_{\rm cool} \approx 2 \cdot 10^6$ yr and $L_{\gamma} \approx 10^{45 - 46} ~{\rm erg} ~{\rm s}^{-1}$, see Fig.10)
should go mainly into heating of the ICM. Note also that this estimate is a lower limit to the energy available
for heating of the ICM since only electrons which produce emission at $E> 100$ {\rm MeV}  are considered. We
notice that the energy budget $E_{\rm merg}\approx E_{\gamma} / \varepsilon  \sim 6 \cdot 10^{60-61}$ erg (we
assume $\varepsilon = 10^{-2}$) is larger than the kinetic energy of the IC gas, which is of the order of
 \be
 E_{kin} \approx {3 \over 2} N_p n kT \sim 2.2 \cdot 10^{59}  ~{\rm erg}~ \bigg({n \over 10^{-4} ~{\rm cm}^{-3}} \bigg)
 \bigg( {T \over 8 ~{\rm keV}} \bigg)
 \ee
(we assume here a sphere of IC gas with particle density $n$, temperature $T$ and total number of particles $N_p =
M/m_p$ where $M=10^{14} M_{\odot}$ is the gas mass of the cluster and $m_p$ is the proton mass). As a consequence,
one should expect that the just forming/merging clusters suggested by KT and TK have quite high temperatures if
most of the merging energy is transformed  into heating of the ICM, as indicated by numerical simulation (see,
e.g., Sarazin 2001 for a review). Specifically, for $n \approx 10^{-4} ~{\rm cm}^{-3}$, an ICM density which is
appropriate to non-virialized clusters, one should expect to have $T \sim 27 - 270$ keV, values which are by far
higher than the temperatures actually observed in relaxed clusters of similar mass and also higher than those of
the forming/merging clusters found in numerical simulations of structure formation (see, e.g., Roettiger et al.
1999, Ricker \& Sarazin 2001, Schindler 2001). This result indicates again that the hypothesis that the EGRET
source luminosity is provided by just forming/merging clusters is somewhat extreme.\\
 Finally, we notice here, following Blasi \& Colafrancesco (1999) and Blasi (2000), that values $L_{\gamma} \sim
 10^{45} {\rm erg} ~{\rm s}^{-1}$ provided by primary electrons would imply a quite high diffuse radio emission
 in the case of usual IC magnetic field values $B \sim 1 ~\mu {\rm G}$, which would have been easily detected in these merging clusters,
 unless very low and unreasonable (see, e.g., Carilli \& Taylor 2001) values, $B \ll 1 ~\mu {\rm G}$, are considered.
Moreover, also strong EUV and hard X-ray excesses would be present in many of the merging clusters selected by KT,
which does not seem the case (see, e.g., Bowyer 2000, Lieu et al. 1999, Colafrancesco 2001a).\\
 So, in conclusion, the suggestion that just forming/merging clusters are the counterparts of the unidentified EGRET
 sources at high galactic latitude seems to face several theoretical problems.

On the observational side, we noticed that none of the clusters selected by KT (and found in our analysis
presented in Sect.3 above) show evidence of strong merging. In fact, strong ICM shocks are expected in merging
clusters and their features can be observed in the cluster X-ray images (see, e.g., Sarazin 2001). Shocks are
irreversible changes to the IC gas in clusters and hence increase the entropy $S\propto ln(T/n^{2/3})$ in the gas.
Thus, one can use X-ray observations to determine the temperature $T$ and the density $n$ of the IC gas and hence
to measure the specific entropy in the gas just before and after the apparent merger shocks seen in the X-ray
images. Since merger shocks produce compression, heating, pressure increase and entropy increase, the
corresponding increase in all of these quantities (and in particular the entropy) can be used to check that
discontinuities are really shocks and not ``cold fronts'' or other contact discontinuities (see, e.g., Sarazin
2001 for a discussion). Markevitch et al. (1999) applied such kind of test to the ASCA temperature maps and ROSAT
images of several clusters. There are clear cases, like the Cyg-A cluster, in which a change in entropy is
observed at the shock front thus confirming the presence of a merger shock. On the other hand, cold fronts with no
entropy change at the discontinuity region have been observed in a number of other clusters including A3667
(Vikhlinin et al. 2001), RXJ1720.1+2638 (Mazzotta et al. 2001) and possibly also A754 and A2163.\\
 The cluster A85 which is considered by KT as a candidate for being a strong merging system triggering the gamma-ray
 emission of the EGRET source 3EGJ0038-0949 is clearly associated, instead, with a cold front (a signature of a possible early stage of
 merging, see Sarazin 2001 and references therein), and not with an ongoing violent merging process.\\
 The two clusters A219 and A2963 associated with 3EGJ0158-3602 have very poor information available (see Sect.3.4) especially at X-ray wavelenghts,
 and there is no evidence of merging ongoing in these clusters.\\
 Also the clusters A1555 and A1558 associated with 3EGJ1234-1318 have poor information in X-rays (see Sect.3.7) and there is no
 evidence of merging ongoing in these clusters.\\
 The clusters A1564 and A1581 fall beyond the $95 \%$ position error contours of the EGRET source 3EGJ1235+0233,
 they have poor information in X-rays and there is no
 evidence of merging ongoing in these clusters.\\
 The cluster A1688 associated with 3EGJ1310-0517 has no relevant information in X-rays (see Sect.3.9)
 and there is no evidence of merging ongoing in this clusters.\\
 The cluster A1758 associated with 3EGJ1337+5029 is a distant, bright X-ray cluster (see Sect.3.17)
 which has a temperature and metallicity structure similar to that of nearby clusters with similar richness
 (Rizza et al. 1998). This cluster has two main clumps with irregular, unrelaxed morphology (Rizza et al. 1998).
 However, the presence of either an ongoing merging or a system consisting of two orbiting cold clumps is demanded to
 more detailed X-ray studies with Chandra and/or XMM.\\
 The cluster A1781 associated with 3EGJ1347+2932 is a bright X-ray cluster with a high radio activity in its
 galaxy population (see Sect.3.15). However, there is no evidence of merging ongoing in this clusters.\\
 To summarize, the available observations do not confirm the presence of ongoing, strong merging in the cluster sample
suggested by KT as the possible counterpart of some unidentified EGRET sources.

Moreover, KT also suggested that the brightest unidentified EGRET source 3EG1835+5918 is the gamma-ray counterpart
of a galaxy cluster which is still uncatalogued and should be one of the ``just forming'' gamma-ray clusters
proposed by these authors. This X-ray cluster is well outside the error ellipse of the EGRET source and ``there is
no reason to suspect that they are related'', according to the analysis performed by Mirabal et al. (2000): also,
there is no evidence of an AGN belonging to this cluster. Other reasons that do not indicate any relation between
the cluster and the EGRET source are the high gamma-ray flux, $F_{P1234}(>100 ~{\rm MeV}) = (60.6 \pm 4.4) \cdot
10^{-8} ~\cm2s$, which is much higher than the typical flux  of the EGRET-cluster associations (see Fig.3), and
the very flat spectral index, $\gamma = 1.69 \pm 0.07$, which is much flatter than those of the EGRET-cluster
associations (see Fig.3). Such high gamma-ray flux and flat spectral index are more typical of an AGN or pulsar
(see Fig.3) being the possible counterpart of this bright EGRET source. These conclusions has been reached also
through an independent analysis of this source by Mirabal \& Halpern (2001) and Reimer et al. (2001).

We conclude, on the basis of the available observational and theoretical evidence, that cluster formation/merging
cannot be responsible for most of the gamma-ray emission observed in the clusters associated with the EGRET
sources listed in Table 1.
 As discussed in our paper, the energy release at gamma-ray energies $E> 100$ ~{\rm MeV} of the EGRET-cluster associations is probably due to a
superposition of diffuse (associated with the active ICM of the cluster) and concentrated (associated with the
active galaxies living within the cluster) gamma-ray emission.

While at the moment we have the first, preliminary evidence for the first gamma-rays coming from galaxy clusters,
their detailed study will have a full bloom with the next generation space-borne (AGILE, GLAST, MEGA) and
ground-based (VERITAS, ARGO, MAGIC) gamma-ray instruments. The next generation gamma-ray telescopes, and
especially the GLAST mission, will have the spatial and spectral capabilities to confirm the preliminary result
here presented and to disentangle between the diffuse and concentrated nature of the cluster gamma-ray emission.\\
 Gamma-ray observations of galaxy clusters in the range $\sim 0.01 - 10^4$ GeV (see Fig.11 for a prediction in
 the case of a Coma-like cluster) can probe directly the existence of
different populations of relativistic particles (e.g., electrons, protons, dark matter particles) in the
intracluster medium through their distinctive gamma-ray spectral features and will open a new window on the
astrophysical studies of large scale structures in the universe.
 Moreover, the detection of mid-energy ($\sim 10 -
100$ {\rm MeV}) and high-energy ($>100$ ~{\rm MeV}) gamma-rays from galaxy clusters will definitely disentangle
the leading mechanisms for the origin of the variety of puzzling non-thermal phenomena (radio halos/relics, EUV
and hard X-ray excesses) which are already observed in many galaxy clusters.

\vskip1.truecm
  {\bf Acknowledgments}

The author (S.C.) thanks the Referee for several useful suggestions which contributed to improve both the clarity
and the presentation of the paper. Part of the data analysis has been performed at the ASI Science Data Center
with the collaboration of P. Giommi. S.C. thanks also G. Ghisellini, M. Salvati, D. Fargion and P. Sreekumar for
useful discussions.
% Correspondence should be addressed to S.C. (e$-$mail: cola@coma.mporzio.astro.it).

%\clearpage
\begin{table}
\begin{center}
\begin{tabular}{|c|c|c|c|c|c|c|c|c|c|c|}
\hline
             &    &     &         &    &     &     &   &           &       \\
EGRET source & RA & DEC & Cluster & RA & DEC & $z$ & R & $r_{opt}$ & Notes \\
             &    &     &         &    &     &     &   &           &       \\
\hline

* 3EG J2219-7941  & 22 20 00.0  & -79 41 24.00 & A1014S &  22 24 10 & -80 10 4  & 0.117 & 2 & 13\arcmin  & NVSS     \\

                &             &              & A1024S &  22 27 32 & -78 45 4    & 0.073 & 1 & 17\arcmin  & RG, NVSS \\

  3EG J1825-7926  & 18 25 02.4  & -79 26 24.00 & A3631  &  18 34 08 & -78 47 4  &   -    & 0 &     -       &   -
\\

  3EG J0348-5708  & 03 48 28.8  & -57 08 24.00 & A3164  &  03 45 49 & -57 02 4  & 0.057 & 0 & 4.5\arcmin &    -
\\

* 3EG J0159-3603 & 01 59 21.6   & -36 03 36.00 & A2963  &  02 00 45 & -35 59 3  &   -    & 0 &    -        & NVSS \\

               &              &              & A219S  &  02 02 03 & -35 48 3    & 0.135  & 1 & 15\arcmin  &  -    \\

  3EG J0616-3310 &  06 16 36.0  & -33 10 12.00 & A577S  &  06 15 18 & -34 07 0  & 0.235  & 2 & 7\arcmin   &    -
\\

               &              &              & A575S  &  06 13 25 & -33 40 5    & 0.135  & 0 &     -       & NVSS \\

               &              &              & A573S  &  06 12 02 & -32 57 4    & 0.204  & 1 & 12\arcmin  & NVSS \\

  3EG J2034-3110 &  20 34 55.2  & -31 10 48.00 & A886S  &  20 37 11 & -31 38 3  & 0.203  & 1 & 12\arcmin  & NVSS
\\

  3EG J1234-1318 &  12 34 02.4  & -13 18 36.00 & A1558  &  12 33 59 & -13 34 3 & 0.145  & 0 & 14\arcmin  &   -
\\

               &              &              & A1555  &  12 31 59 & -13 23 3   & 0.150   & 1 & 14\arcmin  &   -  \\

* 3EG J0038-0949 &  00 38 57.6  & -09 49 12.00 & A85 &  00 41 37  & -09 20 3 & 0.056     & 1 & 30\arcmin  & RH, RG, NVSS \\

* 3EG J1310-0517 &  13 10 24.0  & -05 18 00.00 & A1688  &  13 11 29  & -04 40 5 & 0.190 & 0 &   -         & NVSS \\

* 3EG J0253-0345 &  02 53 57.6  & -03 45 36.00 & A388   &  02 51 36  & -03 45 4 & 0.134 & 2 & 10\arcmin  & NVSS \\

* 3EG J0439+1105 & 04 39 14.4  & 11 05 24.00   & A497   &  04 36 51  & 10 38 0 & 0.140  & 0 & 17.5\arcmin & NVSS \\

* 3EG J0215+1123 &  02 16 00.0 &  11 22 48.00 & A331    & 02 15 35   & 11 21 5 & 0.186  & 1 & 9\arcmin   & NVSS \\

  3EG J2248+1745 &  22 48 57.6 & 17 46 12.00 & A2486    & 22 48 45   & 17 09 5 & 0.143  & 0 & 18\arcmin  & NVSS
\\

  3EG J1212+2304 &  12 12 36.0 &  23 04 48.00 & A1494   & 12 13 14   & 23 56 1 & 0.159  & 1 & 15\arcmin  & NVSS
\\

  3EG J1347+2932 &  13 47 12.0 &  29 32 24.00 & A1781   & 13 44 28   & 29 50 5 & 0.062  & 0 & 16\arcmin  & RG \\

* 3EG J1424+3734 &  14 24 52.8 &  37 34 48.00  & A1902  & 14 21 46   & 37 17 2 & 0.160  & 2 & 15\arcmin  & RG, NVSS \\

               &              &              & A1914  & 14 26 02   & 37 49 3  & 0.171   & 2 & 13\arcmin  & RH, RG, NVSS \\

* 3EG J1337+5029 &  13 37 31.2 & 50 28 48.00 & A1758    & 13 32 32   & 50 30 3  & 0.279 & 3 & 11\arcmin  & RH, RG, NVSS \\

  3EG J1447-3936 &  14 14748.0 & -39 36 36.0 & A774S    & 14 49 23   & -40 20 6 & 0.127 & 0 & 13.5\arcmin & -
\\
\hline
\end{tabular}
\end{center}
\caption{
 {\bf List of probable cluster - EGRET source association}.
 Shown are the coordinates of the EGRET
 unidentified sources (Cols. 2 and 3) together with those of the associated galaxy clusters
 (Cols. 5 and 6). The cluster optical redshifts (Col. 7), their richnesses R (Col. 8) and optical radii $r_{opt}$ (Col.9) are extracted
 from the NED archive. The most probable associations are marked with an asterisk (see text for details).
 Notes: RG: identified radio galaxies in the cluster; NVSS: radio sources found within the Abell radius,
 $3 h^{-1}_{50}$ Mpc, of the cluster; RH: radio halo or relic belonging to the cluster.
 The specific ID names of the identified radio galaxies and NVSS radio sources associated to the clusters
 are given in Sect.3 of the text.
 }
\end{table}


\begin{thebibliography}{}
\bibitem{} Abell, G.O., Corwin, H.G. \& Olowin, R.P. 1989, ApJS, 70, 1
\bibitem{} Arnaud, M. \& Evrard, A., 1999, MNRAS, 305, 631
\bibitem{} Bagchi, J. et al. 1998, MNRAS, 296, L23
\bibitem{} Berezinsky V.S., Bulanov, S.V., Dogiel, V.A., Ginzburg, V.L. \& Ptuskin, V.S.
1990, {\it Astrophysics of Cosmic Rays}, Nort-Holland, Amsterdam
\bibitem{} Blandford, R.D. 2001, in {\it Particles and Fields in Radio Galaxies}, ASP Conference Series, Eds.
R.A.Laing and K.M. Blundell, in press (astro-ph/0110395)
\bibitem{} Blasi, P. 2000, Astroparticle Physics, 15, 223
\bibitem{} Blasi, P. \& Colafrancesco, S. 1999, Astroparticle Physics, 12, 169
\bibitem{} Boehringer, H. et al. 2000, ApJS, 129, 435
\bibitem{} Bowyer, S., 2000, Proceedings of the {\it American Astronomical Society, HEAD
Meeting}, 32, 1707
\bibitem{} Briel, U.G. \& Henry, P. 1993, A\&A, 278, 379
\bibitem{} Brinkman, W. et al. 1995, A\&AS, 109, 147
\bibitem{} Carilli, C.L. \& Taylor, G.B. 2001, ARAA, in press (astro-ph/0110655)
\bibitem{} Colafrancesco S., 2001a, in {\it Constructing the
universe with clusters of galaxies}, Eds. F. Durret and D. Gerbal, (2001)
\bibitem{} Colafrancesco S., 2001b, in {\it Gamma 2001}, Eds. S. Ritz et al. (2001), p.427
\bibitem{} Colafrancesco S. \& Blasi P., 1998, Astroparticle Physics, 9, 227
\bibitem{} Colafranesco S. \& Mele B., 2001, ApJ, 562, 24
\bibitem{} Condon, J.J. et al. 1998, AJ, 115, 1693
\bibitem{} Dar, A. \& DeRujula, A., 2000, {\it CERN-TH/2000-216}, preprint astro-ph/0007306
\bibitem{} Ebeling, H. et al. 1996, MNRAS, 281, 799
\bibitem{} Ebeling, H. et al. 1998, MNRAS, 301, 881
\bibitem{} Feretti, L., Perola, G.C. \& Fanti, R. 1992, A\&A, 265, 9
\bibitem{} Fusco-Femiano R. et al., 1999, ApJ, 513, L21
\bibitem{} Fusco-Femiano R. et al., 2000, ApJ, 534, L7
\bibitem{} Gehrels, N., Macomb, D.J., Bertsch, D.L., Thompson, D.J. \&
 Hartman, R. 2000, Nature, 404, 363
\bibitem{} Giovannini, G. \& Feretti, L.,  2000, New Astronomy, 5, 535
\bibitem{} Gregory, P.L. \& Condon, J.J. 1991, ApJS, 75, 1011
\bibitem{} Grenier, I.A., 2001, in {\it Gamma-Ray 2001}, Eds., Gehrels, Ritz and Shrader:
AIP Conference Proceedings, in press
\bibitem{} Hartman, R.C. et al., 1999, ApJS, 123, 79
\bibitem{} Henriksen, M., 2000, ApJ, 511, 666
\bibitem{} Inoue, S. \& Sasaki, S. 2001, preprint astro-ph/0106187
\bibitem{} Kaastra J. et al., 1999, ApJ, 519, L119
\bibitem{} Kaiser, C.R. \& Alexander, P. 1999, MNRAS, 305, 707
\bibitem{} Kanbach, G. et al., 1988, Space Sci. Rev., 49, 69
\bibitem{} Kawasaki, W. \& Totani, T. 2001, preprint astro-ph/0108309 (KT)
\bibitem{} Kowalski, M.P., Cruddace, R.G., Wood, K.S. \& Ulmer, M.P. 1984, ApJS, 56, 403
\bibitem{} Komissarov, S.S. \& Gubanov, A.G. 1994, A\&A, 285, 27
\bibitem{} Lieu R. et al., 1999, ApJ, 510, L25
\bibitem{} Lima Neto, G.B., Pislar, V. \& Bagchi, J. 2001, A\&A, 368, 440
\bibitem{} Longair, M. 1993, {\it High Energy Astrophysics}, Cambridge University Press, Cambridge
\bibitem{} Markevitch, M., Sarazin, C.L. \& Vikhlinin, A. 1999, ApJ, 521, 526
\bibitem{} Mazzotta, P. et al. 2001, ApJ, in press
\bibitem{} Miniati, F., Ryu, D., Kang, H. \& Jones, T.W. 2001, preprint astro-ph/0105465
\bibitem{} Mirabal, N., Halpern, J.P., Eracleous, M. \& Becker, R.H. 2000, ApJ, 543, 697
\bibitem{} Mirabal, N. \& Halpern, J.P. 2001, ApJ, 547, L137
\bibitem{} Nath, B.B. \& Roychowdhury, S. 2002, preprint astro-ph/0202201
\bibitem{} Owen, F.N. \& Ledlow, M.J. 1997, ApJS, 108, 41
\bibitem{} Padovani, P., Ghisellini, G., Fabian, A.C. \& Celotti, A. 1993, MNRAS, 260, L21
\bibitem{} Reimer, W. et al. 2001, MNRAS, 324, 772
\bibitem{} Rephaeli, Y., Gruber, W. \& Blanco, P., 1999, ApJ, 511, L21
\bibitem{} Ricker, P.M. \& Sarazin, C.L. 2001, ApJ, 561, 621
\bibitem{} Rizza, E. et al. 1998, MNRAS, 301, 328
\bibitem{} Roettiger, K., Burns, J.O. \& Stone, J.M.  1999, ApJ, 518, 594
\bibitem{} Sakelliou, I. \& Merrifield, M.R. 2000, MNRAS, 311, 649
\bibitem{} Sarazin, C.L. 1988 {\it X-Ray Emission from Clusters of Galaxies} (Cambridge: Cambridge Univ.
Press)
\bibitem{} Sarazin, C.L. 2001, in {\it Merging Processes in Clusters of Galaxies}, Eds. L. Feretti, I.Gioia and G.
Giovannini, (Dordrecht: Kluwer), in press (astro-ph/0105418)
\bibitem{} Schindler, S. 2001, in {\it Merging Processes in Clusters of Galaxies}, Eds. L. Feretti, I.Gioia and G.
Giovannini, (Dordrecht: Kluwer), in press
\bibitem{} Sreekumar, P. et al., 1996, ApJ, 464, 628
\bibitem{} Struble, M.F., Mitchell F., \& Rood, H.J. 1999, ApJS, 125, 35
\bibitem{} Totani, T. \& Kitayama, T. 2000, ApJ, 545, 572 (TK)
\bibitem{} Ulmer, M.P. 1980, ApJ, 235, 351
\bibitem{} Ulrich, M.-H., Maraschi, L. \& Urry, M., 1997, ARA\&A, 35, 445
\bibitem{} Urry, M.C. \& Padovani, P. 1995, PASP, 107, 803
\bibitem{} Vikhlinin, A., Markevitch, M. \& Murray, S. 2001, ApJ, 551, 160
\bibitem{} V\" olk H. \& Atoyan, A., 2000, Astroparticle Physics,in press,  preprint astro-ph/9812458
\bibitem{} White, R.L. \& Becker, R.H. 1992, ApJS, 79, 331
\bibitem{} Wu, X.P., Xue, Y.J. \& Fang, L.Z. 1999, ApJ, 524, 22
\bibitem{} Yamada, M. \& Fujita, Y. 2001, preprint astro-ph/0105102

\end{thebibliography}
\end{document}